\documentclass[submission,copyright,creativecommons]{eptcs}
\providecommand{\event}{SYNT 2016} 

\usepackage{underscore}
\usepackage{graphicx}
\usepackage{adjustbox}
\usepackage{array}
\usepackage{booktabs}
\usepackage{caption}    
\usepackage{subcaption} 
\usepackage{multirow}
\usepackage{tikz}

\title{SyGuS-Comp 2016: Results and Analysis}
\author{
Rajeev Alur 
\institute{University of Pennsylvania}
\and
Dana Fisman
\institute{Ben-Gurion University}
\and
Rishabh Singh
\institute{Microsoft Research, Redmond}
\and
Armando Solar-Lezama
\institute{Massachusetts Institute of Technology}
}
\def\titlerunning{SyGuS-Comp 2016: Results and Analysis}
\def\authorrunning{R. Alur, D. Fisman, R. Singh \& A. Solar-Lezama}
\begin{document}
\maketitle

\newcommand{\alc}{\textsc{Alchemist-cs}}
\newcommand{\alccsdt}{\textsc{Alchemist-csdt}}
\newcommand{\cvc}{\textsc{Cvc4-1.5}}
\newcommand{\enum}{\textsc{Enumerative}}
\newcommand{\skac}{\textsc{Sketch-ac}}
\newcommand{\ice}{\textsc{Ice-dt}}
\newcommand{\toast}{\textsc{SosyToast}}
\newcommand{\stoch}{\textsc{Stochastic}}
\newcommand{\cvcnew}{\textsc{Cvc4-1.5.1}}
\newcommand{\eusolver}{\textsc{EUSolver}}
\newcommand{\sygus}{SyGuS}
\newcommand{\comp}{SyGuS-Comp}

\begin{abstract}
\emph{Syntax-Guided Synthesis (SyGuS)} is
the computational problem of finding an implementation $f$ that
meets both a semantic constraint
given by a logical formula $\varphi$ in a background theory $T$,
and a syntactic constraint given by a grammar $G$, which specifies the allowed set of
candidate implementations.
Such a synthesis problem can be formally defined in SyGuS-IF,
a language that is built on top of SMT-LIB.

The \emph{Syntax-Guided Synthesis Competition (\comp)} is an
effort to facilitate, bring together and accelerate research and development of efficient
solvers for SyGuS by providing a platform for evaluating different synthesis
techniques on a comprehensive set of benchmarks. 
In this year's competition we added a new track devoted to \emph{programming by examples}. 
This track consisted of two categories, one using the theory of bit-vectors and one using the theory of strings.   
This paper presents and analyses the results of \comp'16.
\end{abstract}

\section{Introduction}
\label{sec:intro}

The Syntax-Guided Synthesis Competition (SyGuS-Comp) was originally developed as a community effort in order to provide an objective basis to compare different approaches to solving the Syntax-Guided Synthesis problem.  In this style of synthesis, the user provides a specification in the form of a logical formula $\varphi$ in a background theory $T$, and a space of programs given as a grammar G; the goal of the synthesizer is to find a program in the space that satisfies the given specification. Concretely, if the specification uses an unknown function $f$, the goal is to find a function $f_{imp}$ that is expressible in the grammar $G$ and such that the formula $\varphi[f/f_{imp}]$ is valid for all values of its free variables.

One of the achievements of the community effort behind SyGuS-Comp has been the development of a standard format for benchmarks.  The \sygus{} format is detailed in other publications~\cite{SyGuSComp15,AlurBDF0JKMMRSSSSTU15},\cite{RaghothamanU14,SyGuS15format,SyGuS16format}, but at a high-level, the \sygus{} format is based on the popular SMT-LIB format for defining SMT problems and is extended to support the description of function grammars. The \sygus{} format has been extended over the last two years to provide special support for important classes of problems such as Invariant Synthesis, or problems involving expressions in integer linear arithmetic~\cite{SyGuS15format}. In its 2016 iteration, the format was also extended to support Programming by Example problems~\cite{SyGuS16format}, which are becoming an important area of study in the synthesis community. 

In the short time that the formalism has been in public circulation, it has already performed well in its goal of facilitating research in synthesis while providing a basis for objective comparison of different algorithms. For example, the competition has provided important insights into the relative merits of different algorithms~\cite{AlurBJMRSSSTU13,AlurBDF0JKMMRSSSSTU15,SyGuSComp15} which have been exploited to help develop and evaluate new algorithms~\cite{GLMN14,JeonQSF15,ReynoldsDKTB15, Saha0M15, AleksICSE15, AlurCAV15,GNMR16}. Beyond synthesizer developers, there is a growing community of users that is coalescing around the formalism. 

SyGuS has found various interesting applications among which are motion planning~\cite{CN16}, compiler optimizations, and cybersecurity~\cite{EldibWW16}. Remarkably, Eldib et al. report that a circuit for mitigating time-delay attacks generated via SyGuS is much smaller than a handcrafted circuit mitigating the same attack, as well as the original circuit (which is vulnerable to that attack). The SyGuS generated circuit used 13 gates compared to 41 of the handcrafted circuit and 21 of the original, and has a shorter critical path: 3 unit delay vs. 6 unit delay of the mitigated and original circuits.

\subsection{The General Track}
We now illustrate the key ideas behind the main formalism and the extensions that have been added in the last two years through a series of illustrative examples. 
\paragraph{Example}
We illustrate the general SyGuS-Comp formalism with this simple example from one of the competition benchmarks. This example is taken from an implementation of a quantum control computer (QCC).\footnote{We thank Nissim Ofek (Yale) for contributing these benchmarks.} The QCC uses expressions from the following grammar:
$$g ::= c ~|~ g + g ~|~ g - g ~|~g ? g$$
where $c$ is any integer constant, $+$ is addition, $-$ is substraction, and $a? b$ stands for 
 ``if $a \geq 0$ then $a$ else $b$''. 
This minimal set of instruction is used to enable a fast implementation. High level commands should be translated to this grammar using a minimal number of operations, since these operations participate in a pipeline, thus every unnecessary delay multiplies. The goal in the following benchmark is to find two functions \texttt{qm-inner-loop} and \texttt{qm-outer-loop} that decrement an inner and outer loop (the inner from 7 to 0, the outer from 3 to 0) formally defined as follows for $x \geq 0$ and $y \geq 0$.

\[
\mbox{\texttt{qm-inner-loop}}(x)= 
\left \{
\begin{array}{ll}
7 & \mbox{if } x=0  \\
x-1 & \mbox{if } \mbox{otherwise} \\
\end{array}
\right .
\]

\[
\mbox{\texttt{qm-outer-loop}}(x,y)= 
\left \{
\begin{array}{ll}
3 & \mbox{if } x=0 \ \wedge\ y=0  \\
y-1 & \mbox{if } x=0\ \wedge\ y\neq 0 \\
y & \mbox{otherwise} 
\end{array}
\right .
\]

These constraints can be succinctly expressed in the \sygus{} format as shown below.
\begin{verbatim}
(set-logic LIA)

(define-fun qm ((a Int) (b Int)) Int
               (ite (< a 0) b a))

(synth-fun qm-inner-loop ((x Int)) Int
   ((Start Int (x
                0
                1
                7
               (- Start Start)
               (+ Start Start)
               (qm Start Start)))))

(synth-fun qm-outer-loop ((x Int) (y Int)) Int
   ((Start Int (x
                y
                0
                1
                3
                (- Start Start)
                (+ Start Start)
                (qm Start Start)))))



(declare-var x Int)
(declare-var y Int)

(constraint (or (< x 0)) 
            (= (qm-inner-loop x)  
               (ite (= x 0) 7 (- x 1))))


(constraint (or (or (< x 0) (< y 0)) 
            (= (qm-outer-loop x y)  
               (ite (= x 0) (ite (= y 0) 3 (- y 1)) 
                             y))))

(check-synth)
\end{verbatim}
\vspace{2mm}
The \texttt{define-fun} command provides the description of the `?' or \texttt{qm} primitive function: 
\[
\mbox{\texttt{qm}}(a,b) = \left \{ 
\begin{array}{ll}
a & \mbox{if } a\geq 0  \\
b & \mbox{otherwise } 
\end{array}
\right .
\]
The \texttt{set-logic} directive indicates that the constraints should be interpreted in terms of the theory of linear integer arithmetic. The directive \texttt{declare-var} is used to declare \texttt{x} and \texttt{y} as universally quantified integer variables. The constraints are introduced with the directive \texttt{constraint}, and \texttt{check-synth} marks the end of the problem and prompts the synthesizer to solve for the missing function.
Crucially, in order for the synthesizer to generate \texttt{qm-inner-loop} and \texttt{qm-outer-loop}, it needs a grammar, which is provided as part of the \texttt{synth-fun} directive. The specified grammar provides exactly the set of allowed operations for the QCC.

\subsection{Conditional Linear Integer Arithmetic Track}
For problems where the grammar consists of the set of all possible integer linear arithmetic terms, it is sometimes possible to apply specialized solution techniques that exploit the information that decision procedures for integer linear arithmetic are able to produce. The 2015 \sygus{} competition included a separate track where the grammar for all the unknown functions was assumed to be the entire theory of Integer Linear Arithmetic with ITE conditionals.

\paragraph{Example} 
As a simple example, consider the problem of synthesizing a function \texttt{abs} that produces the absolute value of an integer. The problem can be specified with the constraint below:
\begin{verbatim}
(set-logic LIA)
(synth-fun abs ((x Int)) Int)
(declare-var x Int)
(constraint (>= (abs x) 0))
(constraint (or (= x (abs x)) (or (= (- x) (abs x)))))
(check-synth)
\end{verbatim}

Note that the definition of the unknown function \texttt{abs} does not include a grammar this time, but because the problem is defined in the theory of linear integer arithmetic (\texttt{LIA}), the default grammar consists of all the operations available in the theory.

\subsection{Invariant Synthesis Track}
One of the main applications of \sygus{} is invariant synthesis. For this problem, the goal is to discover an invariant that makes the verification condition for a given loop valid. Such a problem can be easily encoded in \sygus{}, but invariant synthesis problems have structure that some solution algorithms are able to exploit and that can be lost when encoding it into \sygus{}. Like the 2015 competition, the 2016 competition also included a separate track for invariant synthesis problems where the additional structure is made apparent. In the invariant synthesis version of the \sygus\ format, the constraints are separated into pre-condition, post-condition and transition relation, and the grammar for the unknown invariant is assumed to be the same as that for the conditional linear arithmetic track. We illustrate this format with an example from last year's report~\cite{SyGuSComp15}.

\paragraph{Example} 
For example, consider the following simple loop. 
\begin{verbatim}
Pre: i >= 0  and  j=j0 and i=i0;
while(i > 0){
    i = i - 1;
    j = j + 1;
}
Post: j = j0 + i0;
\end{verbatim}
Suppose we want to prove that the value of j at the end of the loop equals the value of i + j at the beginning of the loop. The verification condition for this loop would check that (a) the precondition implies the invariant, (b) that the invariant is inductive, so if it holds before an iteration and the loop condition is true, then it will hold after that iteration, and (c) that the invariant together with the negation of the loop condition implies the postcondition. All of these constraints can be expressed in the standard \sygus{} format, but they can be expressed more concisely using the extensions explicitly defined for this purpose. Specifically, the encoding will be as follows.
\begin{verbatim}
(set-logic LIA)

(synth-inv inv-f ((i Int) (j Int) (i0 Int) (j0 Int)))

(declare-primed-var i0 Int)
(declare-primed-var j0 Int)
(declare-primed-var i  Int)
(declare-primed-var j  Int)

(define-fun pre-f ((i Int) (j Int) (i0 Int) (j0 Int)) Bool
                  (and (>= i 0) (and (= i i0) (= j j0))))

(define-fun trans-f ((i Int) (j Int) (i0 Int) (j0 Int)
                     (i! Int) (j! Int) (i0! Int) (j0! Int)) Bool
                     (and (and (= i! (- i 1)) (= j! (+ j 1)))
                          (and (= i0! i0) (= j0! j0))))

(define-fun post-f ((i Int) (j Int) (i0 Int) (j0 Int)) Bool
                   (= j (+ j0 i0)))

(inv-constraint inv-f pre-f trans-f post-f)

(check-synth)
\end{verbatim}

The directive \texttt{(declare-primed-var i)} is equivalent to separately declaring \texttt{i} and \texttt{i!}, where the primed version of the variables is used to distinguish their value before and after the loop body. Just like in the earlier example, the function to be synthesized \texttt{inv\_f} does not include a grammar, so the entire \texttt{LIA} grammar is assumed. The constraint \texttt{inv-constraint} is syntactic sugar for the full verification condition involving the invariant, precondition, postcondition and transition function.

\subsection{Programming By Example Track}
There has been a lot of recent interest in the synthesis community for learning programs from examples. Programming By Examples (PBE) systems have been developed for many domains including string transformations~\cite{flashfill,cacm12,blinkfill}, data structure manipulations~\cite{storyboard1,storyboard2}, interactive parser synthesis~\cite{parsersynthesis}, higher-order functional programs over recursive data types~\cite{types1,types2}, and program refactorings~\cite{refactoring}. The 2016 competition included a new separate track for Programming by Examples. The grammar for benchmarks in this track is specified using a context-free grammar similar to the general SyGuS track, but the specification constraints can only be specified using input-output examples. The benchmarks in this track included theory of integers, bit-vectors, and strings.

\paragraph{Example}

Consider the following task taken from FlashFill~\cite{flashfill,cacm12} that requires learning a string transformation program that constructs the initials of the first and last names.

\begin{table}[b]
	\begin{center}
	\begin{tabular}{|c|c|}
	\hline
	{\bf Input} & {\bf Output} \\ \hline \hline
	Nancy FreeHafer & N.F.  \\ \hline
	Andrew Cencici & A.C. \\ \hline
	Jan Kotas & J.K. \\ \hline
	Mariya Sergienko & M.S. \\ \hline
	\end{tabular}
	\end{center}
	\label{pbe-example}
	\caption{a FlashFill example task}
\end{table}

\newpage
The encoding of this problem in the PBE track is as follows:

\begin{verbatim}
(set-logic SLIA)

(synth-fun f ((name String)) String
    ((Start String (ntString))
     (ntString String (name " " "."
                       (str.++ ntString ntString)
                       (str.replace ntString ntString ntString)
                       (str.at ntString ntInt)
                       (int.to.str ntInt)
                       (str.substr ntString ntInt ntInt)))
      (ntInt Int (0 1 2
                  (+ ntInt ntInt)
                  (- ntInt ntInt)
                  (str.len ntString)
                  (str.to.int ntString)
                  (str.indexof ntString ntString ntInt)))
      (ntBool Bool (true false
                    (str.prefixof ntString ntString)
                    (str.suffixof ntString ntString)
                    (str.contains ntString ntString)))))

(declare-var name String)

(constraint (= (f "Nancy FreeHafer") "N.F."))
(constraint (= (f "Andrew Cencici") "A.C."))
(constraint (= (f "Jan Kotas") "J.K."))
(constraint (= (f "Mariya Sergienko") "M.S."))

(check-synth)
\end{verbatim}

The benchmark uses SMT-LIB's SLIA theory that encodes several string functions such as \texttt{str.len}, \texttt{str.indexof}, \texttt{str.contains} etc. All the constant strings that are needed to perform the transformation are also provided as part of the grammar.

\subsection{\comp{}'14 summary}
The first \sygus\ competition, \comp'14\ consisted of a single track --- the general track --- in which the benchmark provides the grammar describing the desired syntactic restrictions for that benchmark. The background theory could be either linear interger arithmetic or bitvectors. Five solvers competed in \comp'14. The solver who won the first place was the \enum\ solver which solved 126 out of 241 benchmarks.

\subsection{\comp{}'15 summary}

The 2015 instance of  \comp{} was the second iteration of the competition and the first iteration to include the separate conditional linear integer arithmetic and invariant synthesis tracks. There were a total of eight solvers submitted to the competition which represented a range of solution strategies. The CVC4-1.5 solver won the general track and the conditional linear integer arithmetic tracks, whereas the ICE DT solver won the invariant synthesis track.

\subsection{\comp{}'16 summary}

The 2016 instance of \comp{} was the third iteration of the competition and included an additional track on Programming By Examples (PBE). In addition to the previous solvers, there were two new solver submitted this year: CVC4-1.5.1 and EUSolver. In the rest of the paper, we describe the details of the benchmarks, new solver strategies, and the results of the competition on different benchmark categories.

\section{Competition Settings}
\label{sec:setting}

\subsection{Participating Benchmarks}
In addition to last year's competition benchmarks, we had $858$ new benchmarks for the Programming By Example (PBE) track. For other tracks, we had the same benchmarks as of last year: General Track (309), CLIA Track (73), and Invariant Synthesis Track (67).

The benchmarks in the PBE track can be classified into two categories:
\begin{itemize}
\item {\bf String Transformations:} The 108 string transformation tasks are taken from public benchmarks of FlashFill~\cite{flashfill,cacm12} and BlinkFill~\cite{blinkfill}. The transformations are defined using a Domain-specific language of string transformations that involve concatenation of substrings of input strings and constant strings, where the substring expressions involve learning positions corresponding to $k^\texttt{th}$ occurrence of a constant string in the inputs. 
\item {\bf Bitvector Transformations:} The 450 bitvector transformation benchmarks were obtained from the 2013 ICFP Programming Competition\footnote{http://icfpc2013.cloudapp.net/}~\cite{icfp}. The programs for the benchmarks were sampled from a bitvector DSL using a strategy of construction k-nuggets (programs of size k that are minimal) and then composing them to generate larger programs. An additional 300 bitvector benchmarks using the same grammar were submitted by Arjun Radhakrishna.
\end{itemize}

\paragraph{String Benchmarks}
The string benchmarks were taken from the public string transformation benchmarks in FlashFill and BlinkFill. These benchmarks correspond to common data cleaning tasks faced by spreadsheet users. The hypothesis space of possible transformations is defined by a DSL that is expressive enough to encode majority of common tasks but at the same time amenable for efficient learning. A subset of the DSL that was encoded in SyGuS benchmark is shown in Figure~\ref{stringdsl}. Note that the SyGuS grammar for these benchmarks currently does not contain loops (Kleene star) and regular expression based position expressions.

\begin{figure}
\begin{verbatim}
(synth-fun f ((x String) (y String)) String
    ((Start String (ntString))
     (ntString String (x y c1 c2 ...
                    (str.++ ntString ntString)
                    (str.replace ntString ntString ntString)
                    (str.at ntString ntInt)
                    (int.to.str ntInt)
                    (ite ntBool Start Start)
                    (str.substr ntString ntInt ntInt)))
      (ntInt Int (0 1 2 (+ ntInt ntInt) (- ntInt ntInt)
                    (str.len ntString) (str.to.int ntString)
                    (str.indexof ntString ntString ntInt)))
      (ntBool Bool (true false
                    (str.prefixof ntString ntString)
                    (str.suffixof ntString ntString)
                    (str.contains ntString ntString)))))

\end{verbatim}
\caption{The grammar for string transformation benchmarks in the PBE track.}
\label{stringdsl}
\end{figure}

The grammar at the top-level consists of string concatenation (\texttt{str.++}) expressions involving constant strings and substring expressions. The constant strings needed for each benchmark are also provided in each benchmark (\texttt{c1}, \texttt{c2}, etc.). For some of the string transformation benchmarks, we created two additional class of benchmarks with the suffix \texttt{-long} and \texttt{-repeat}. The \texttt{-long} benchmarks had 100 input-output examples, whereas the \texttt{-repeat} benchmarks consisted of several input-output examples that were repeated in the constraint. The goal of these additional benchmark categories was to see how increasing the number of examples affects the solver performance, and if solving algorithms can avoid reasoning about repeated input-output examples.

\paragraph{Bitvector Benchmarks}

The bitvector benchmarks were taken from the 2013 ICFP programming contest and the DSL encoded as a SyGuS grammar for the benchmarks is shown in Figure~\ref{bitvectordsl}. Similar to the string transformation DSL, the constants needed for the desired transformation are provided in the grammar.

\begin{figure}
\begin{verbatim}
(define-fun shr1 ((x (BitVec 64))) (BitVec 64) (bvlshr x #x0000000000000001))
(define-fun shr4 ((x (BitVec 64))) (BitVec 64) (bvlshr x #x0000000000000004))
(define-fun shr16 ((x (BitVec 64))) (BitVec 64) (bvlshr x #x0000000000000010))
(define-fun shl1 ((x (BitVec 64))) (BitVec 64) (bvshl x #x0000000000000001))
(define-fun if0 ((x (BitVec 64)) (y (BitVec 64)) (z (BitVec 64))) (BitVec 64) 
  (ite (= x #x0000000000000001) y z))

(synth-fun f ( (x (BitVec 64))) (BitVec 64)
  ((Start (BitVec 64) (c1 c2 ...  x (bvnot Start)
       (shl1 Start) (shr1 Start)
       (shr4 Start) (shr16 Start)
       (bvand Start Start) (bvor Start Start)
       (bvxor Start Start) (bvadd Start Start)
       (if0 Start Start Start)))))
\end{verbatim}
\caption{The grammar for bitvector synthesis benchmarks in the PBE track.}
\label{bitvectordsl}
\end{figure}

The benchmarks for this category were generated from the DSL by first sampling k-nuggest from the DSL and then composing them to obtain larger programs. A k-nugget is a program expression in the DSL of size k such that no other expression in the DSL of size less than k is equivalent with it. The idea in using the k-nuggets for program generation is that the composed programs would lead to more challenging programs that will be less likely to be solved by synthesizing a small equivalent program in the DSL.

\subsection{Participating Solvers}
\label{subsec:solvers}
In addition to 7 solvers from last year's competition, we had two new solver submissions for the 2016 competition: i) CVC4 1.5.1 and ii) EUSolver. Table~\ref{tbl:solvers-in-tracks} summarizes which solver participated in which track.  The two new solvers participated in all 4 tracks.  A total of 6 solvers participated in the General track, 5 in the invariant synthesis track, and 5 in the Conditional Linear arithmetic track. Figure~\ref{tbl:solvers-in-tracks} lists the submitted solvers together with their authors.

The $\cvcnew$ solver employs a refutation-based synthesis approach~\cite{ReynoldsDKTB15}. Instead of solving an exists-forall synthesis formula, it first negates the formula to obtain a forall-exists problem and tries to show it is unsatisfiable. It eliminates the forall quantification over unknown function in two ways: i) if the function is always called with the same parameters in the formula, it skolemizes it with a first-order variable (single invocation case), ii) otherwise if the single invocation property does not hold in the formula, it uses a syntax-guided approach for restricting the space of functions using the grammar. This year's submission had new improvements in both of these two ways. For the single invocation case, the solver has a termination guarantee for LIA and supports newer ways to recognize when a property can be rewritten as single invocation. For the syntax-guided case, it supports an improved symmetry breaking and adds optimizations for unfolding of evaluation functions. For the invariant track, it fixes templates for unknown invariants to improve scalability of the solving algorithm.

The $\eusolver$ combines enumeration with unification to learn complex functions from a grammar that satisfy the specification. It first learns small terms from the function grammar using enumeration such that the learnt terms cover the set of all points. It then synthesizes larger expressions by enumerating predicates and combining them with the learnt terms using a decision tree learning algorithm. It supports multiple sophisticated algorithms for term generation, predicate generation, and unification to compose larger expressions for different categories of benchmarks.

\newcolumntype{R}[2]{%
    >{\adjustbox{angle=#1,lap=\width-(#2)}\bgroup}%
    l%
    <{\egroup}%
}
\newcommand*\rot{\multicolumn{1}{R{90}{1em}}}

\begin{table}[t]
\begin{center}
\begin{tabular}{r||rrrrrrrrr}
 & \multicolumn{9}{c}{Solvers} \\
 Tracks & \rot{\alc} & \rot{\alccsdt} & \rot{\cvc} & \rot{\enum} & \rot{\ice} & \rot{\skac} & \rot{\stoch} & \rot{\cvcnew} & \rot{\eusolver} \\ \hline \hline
 LIA & 1 & 1 & 1 & 0 & 0 & 0 & 0 & 1 & 1\\
 INV & 1 & 0 & 1 & 0 & 1 & 0 & 0 & 1 & 1\\
 General &  0 & 0 & 1 & 1 & 0 & 1 & 1 & 1 & 1\\ 
 PBE &  0 & 0 & 1 & 1 & 0 & 1 & 1 & 1 & 1\\ 
\end{tabular}
\end{center}
\caption{Solvers participating in each track}
\label{tbl:solvers-in-tracks}
\end{table}

\begin{figure}
{\small{
\begin{center}
\begin{tabular}{r||l}
 Solver &  Authors \\ \hline \hline
 \alc & Daniel Neider (UIUC),
        Shambwaditya Saha (UIUC) and
        P. Madhusudan (UIUC) \\
 \alccsdt & Shambwaditya Saha (UIUC),
          Daniel Neider (UIUC) and
          P. Madhusudan (UIUC) \\
 \cvc & Andrew Reynolds (EPFL),
      Viktor Kuncak (EPFL), 
       Cesare Tinelli (Univ. of Iowa), \\ 
       & Clark Barrett (NYU), 
       Morgan Deters (NYU) and
       Tim King (Verimag) \\
 \enum & Abhishek Udupa (Penn) \\
 \ice & Daniel Neider (UIUC), 
       P. Madhusudan (UIUC) and
       Pranav Garg (UIUC) \\
 \skac  & Jinseong Jeon (UMD),	Xiaokang Qiu (MIT),	Armando Solar-Lezama (MIT) and\\
       & 	Jeffrey S. Foster (UMD)	\\
 \stoch & Mukund Raghothama (Penn) \\              
 \cvcnew & Andrew Reynolds (Univ. Of Iowa),
       Cesare Tinelli (Univ. of Iowa), \\ 
       & Clark Barrett (NYU), and
       Tim King (Google) \\
 \eusolver  & Arjun Radhakrishna (Penn) and
          Abhishek Udupa (Microsoft) \\
\end{tabular}
\end{center}
\caption{Submitted solvers and their authors}
\label{tbl:solvers-authors}
}}
\end{figure}

\subsection{Experimental Setup}
\label{subsec:tech}
The solvers were run on the StarExec platform~\cite{starexec} with a dedicated cluster of 12 nodes, where each node consisted of two 4-core 2.4GHz Intel processors with 256GB RAM and a 1TB hard drive. The memory usage limit of each solver run was set to 128GB. The wallclock time unit was set to 3600 seconds (thus, a solver that used all cores could consume at most 14400 seconds cpu time).
 
The solution that the solvers produce are being checked for both syntactic and semantic correctness. That is, a first post-processor checks that the produced expression adheres to the grammar specified in the given benchmark, and if this check passes, a second post-processor checks that the solution adheres to semantic constraints given in the benchmark (by invoking an SMT solver).

\section{Competition Results and Analysis}
\label{sec:results}

\subsection{Results Overview} 
\label{subsec:bench-res-overview}

\begin{figure}
	\centering
	\includegraphics[scale=0.40,width=0.6\textwidth,bb=0 0 415 248]{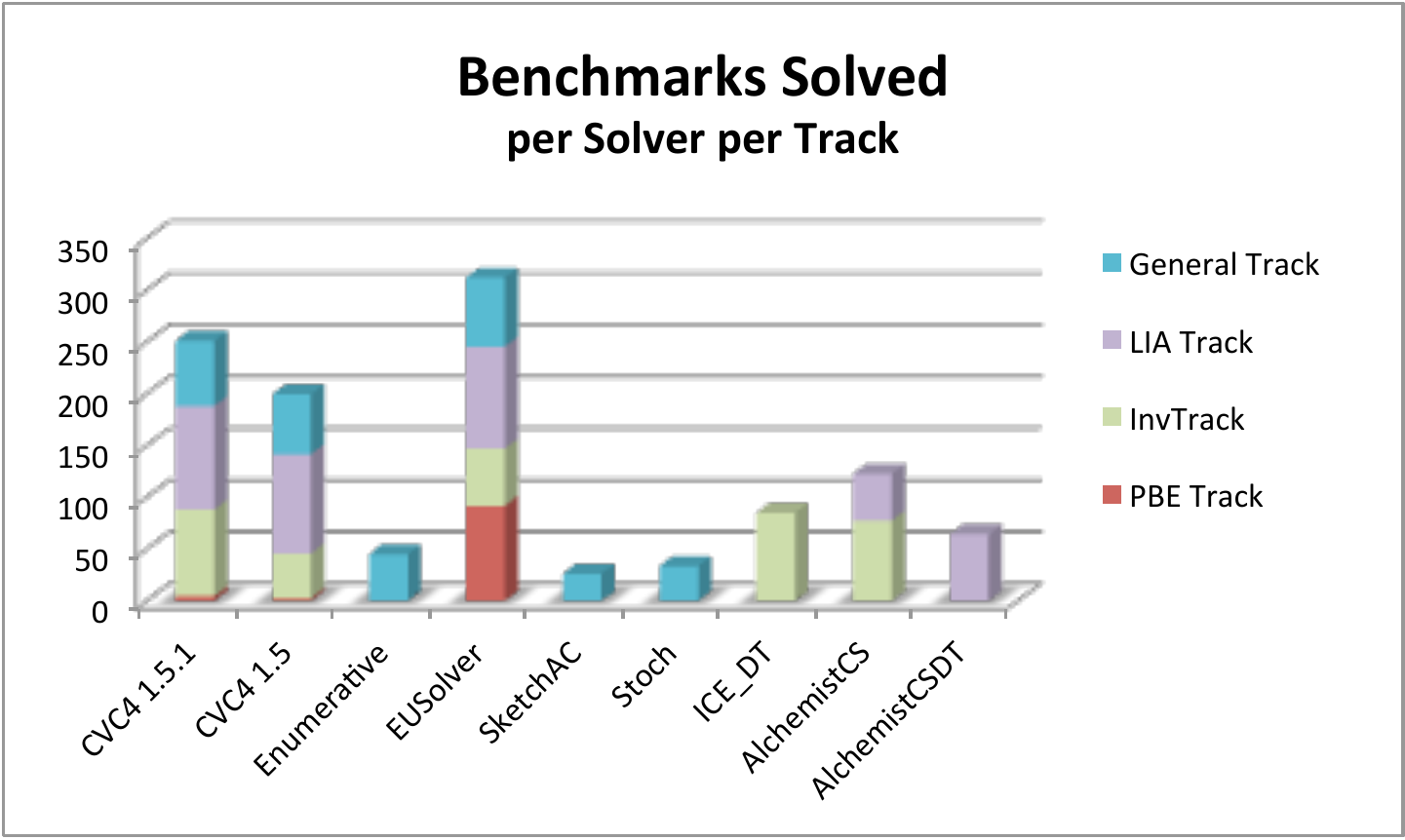}
	\caption{The overall combined results for each solver on benchmarks from all the 4 tracks.}
	\label{combinedresults}
\end{figure}

The combined results for all tracks for each benchmark is shown in Figure~\ref{combinedresults}. The figure shows the sum of percentages of benchmarks solved by the solvers for each category. We can observe that the $\eusolver$ solves the highest percentage of benchmarks in the combined tracks, whereas the $\cvcnew$ solver solves the second most percentage of benchmarks.

The primary criterion for winning a track was the number of benchmarks solved, but we also analyzed the time to solve and the the size of the generated expressions. Both where classified using a pseudo-logarithmic scale as follows.
For time to solve the scale is [0,1), [1,3), [3,10), [10,30),[30, 100), [100,300), [300, 1000), [1000,3600), $>$3600. That is the first ``bucket'' refers to termination in less than one second, the second to termination in one to three second and so on. We say that a solver solved a certain benchmark \emph{among the fastest} if the time it took to solve that benchmark was on the same bucket as that of the solver who solved that benchmark the fastest. 
For the expression sizes the pseudo-logarithmic scale we used is [1,10), [10,30), [30,100), [100,300), [300,1000), $>$1000 where expression size is the number of nodes in the SyGuS parse-tree.
We also report on the number of benchmarks \emph{solved uniquely} by a solver (meaning the number of benchmark that solver was the single solver that managed to solve them).

\paragraph{General Track}

\begin{figure}[t]
	\begin{center}
	\begin{minipage}{.45\textwidth}
		\centering
		\includegraphics[scale=0.6,width=0.98\textwidth,bb=0 0 373 219]{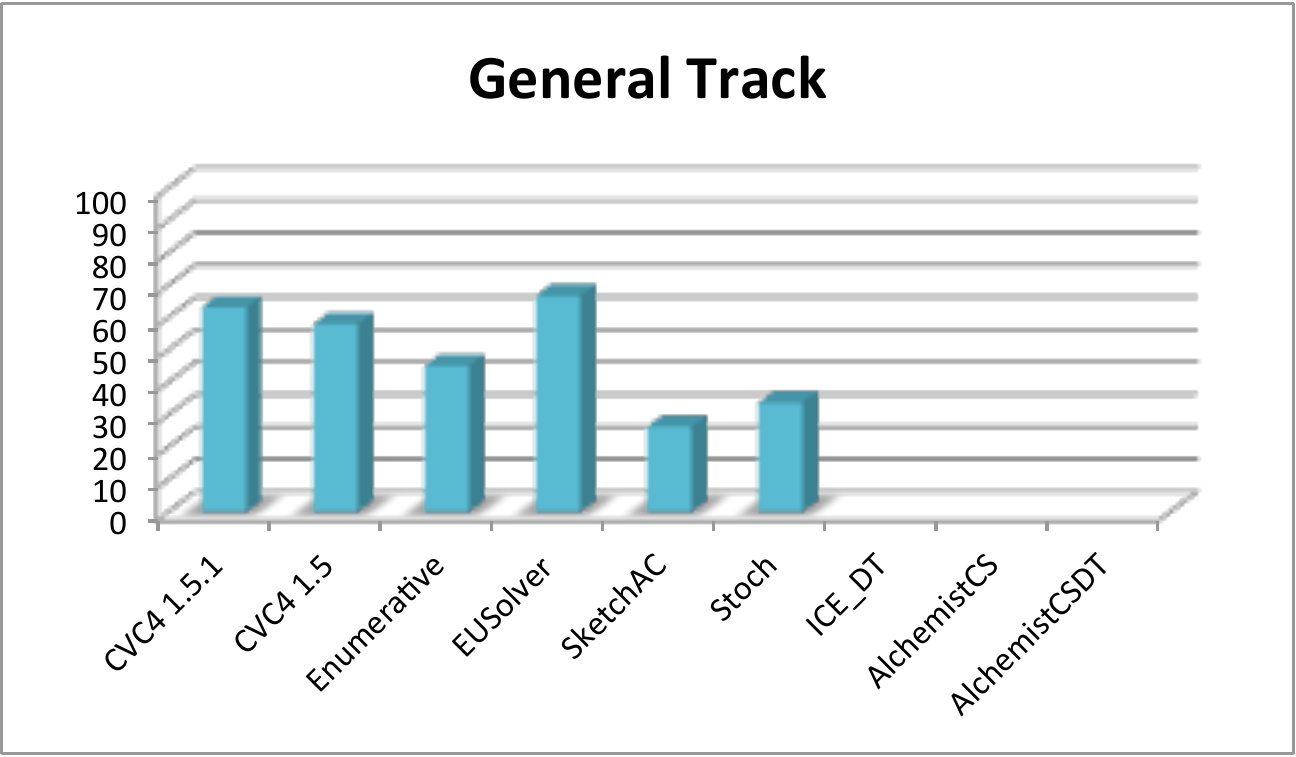}
		\label{fig:generaloverall}
	\end{minipage}
	\begin{minipage}{.45\textwidth}
		\centering
		\includegraphics[scale=0.6,width=0.98\textwidth,bb=0 0 373 219]{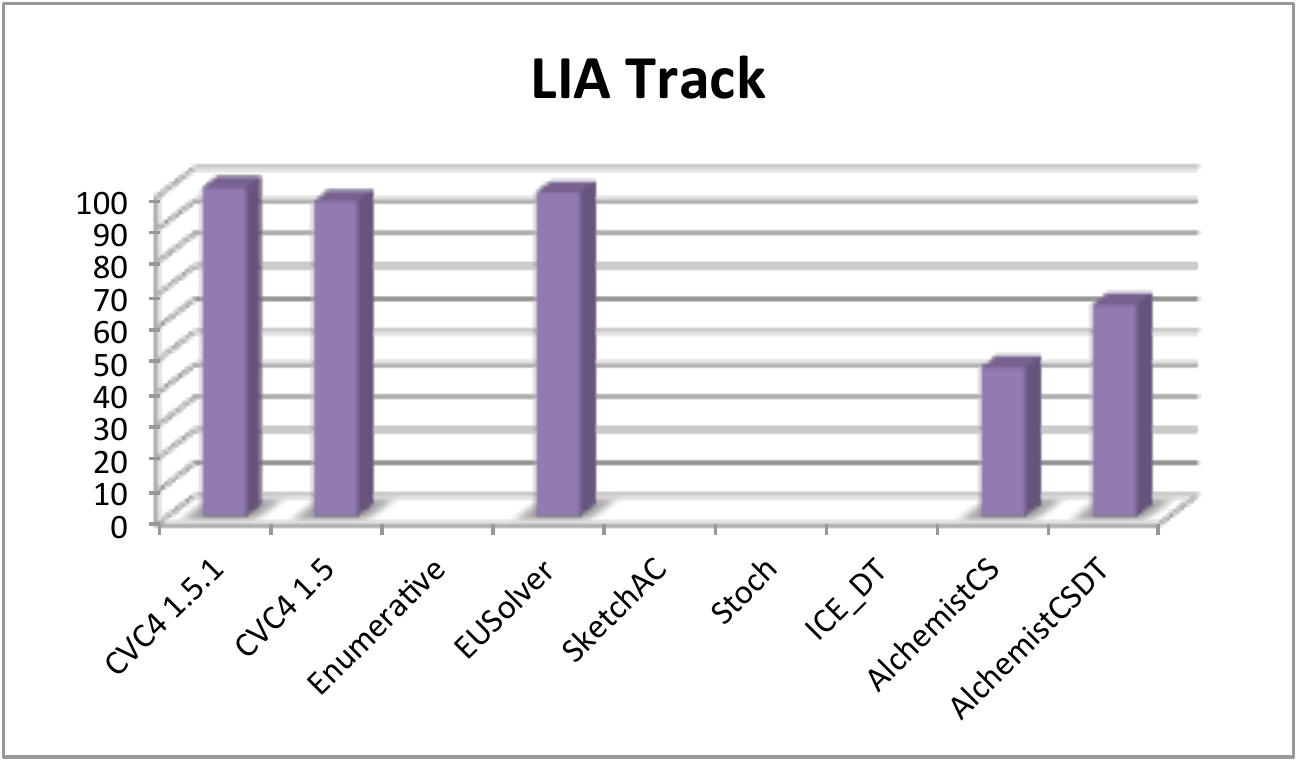}
		\label{fig:liaoverall}
	\end{minipage}
	\\
	\vspace{2mm}
	\begin{minipage}{.45\textwidth}
		\centering
		\includegraphics[scale=0.6,width=0.98\textwidth,bb=0 0 373 219]{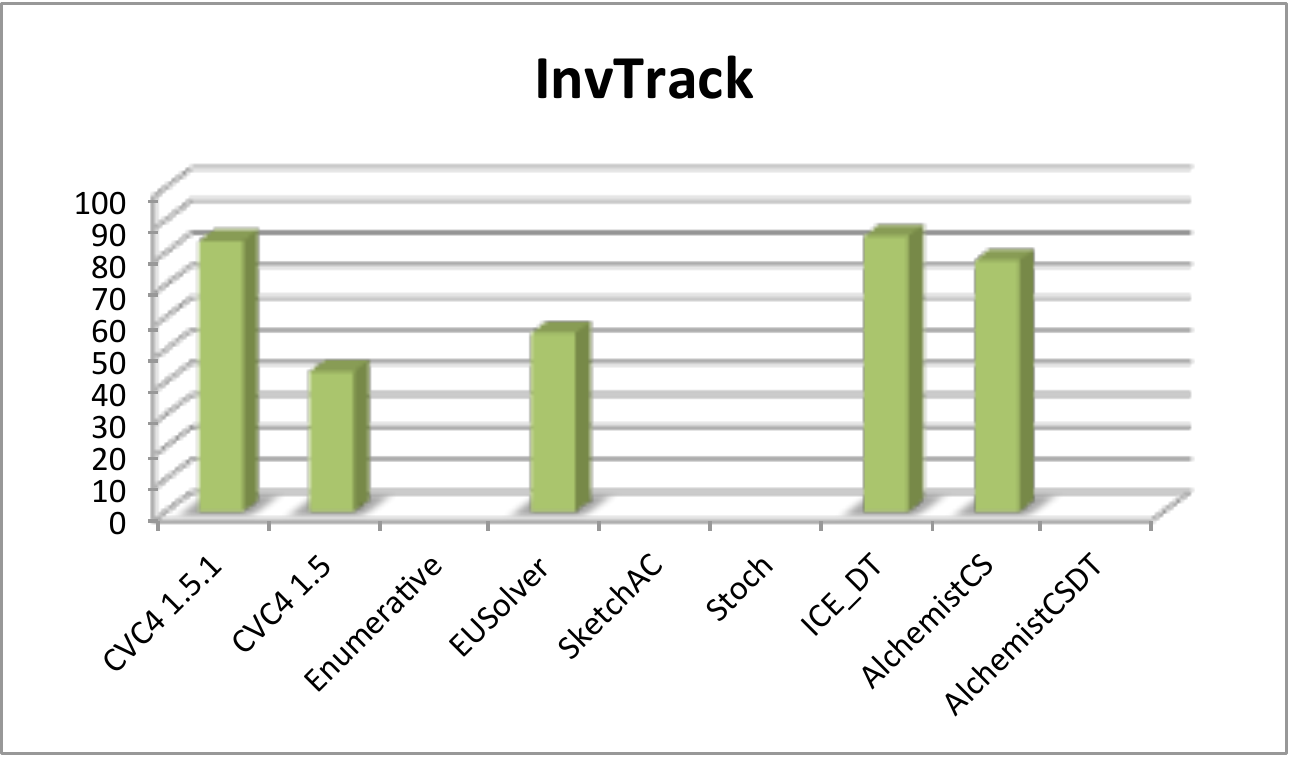}
		\label{fig:invoverall}
	\end{minipage}
	\begin{minipage}{.45\textwidth}
		\centering
		\includegraphics[scale=0.6,width=0.98\textwidth,bb=0 0 373 219]{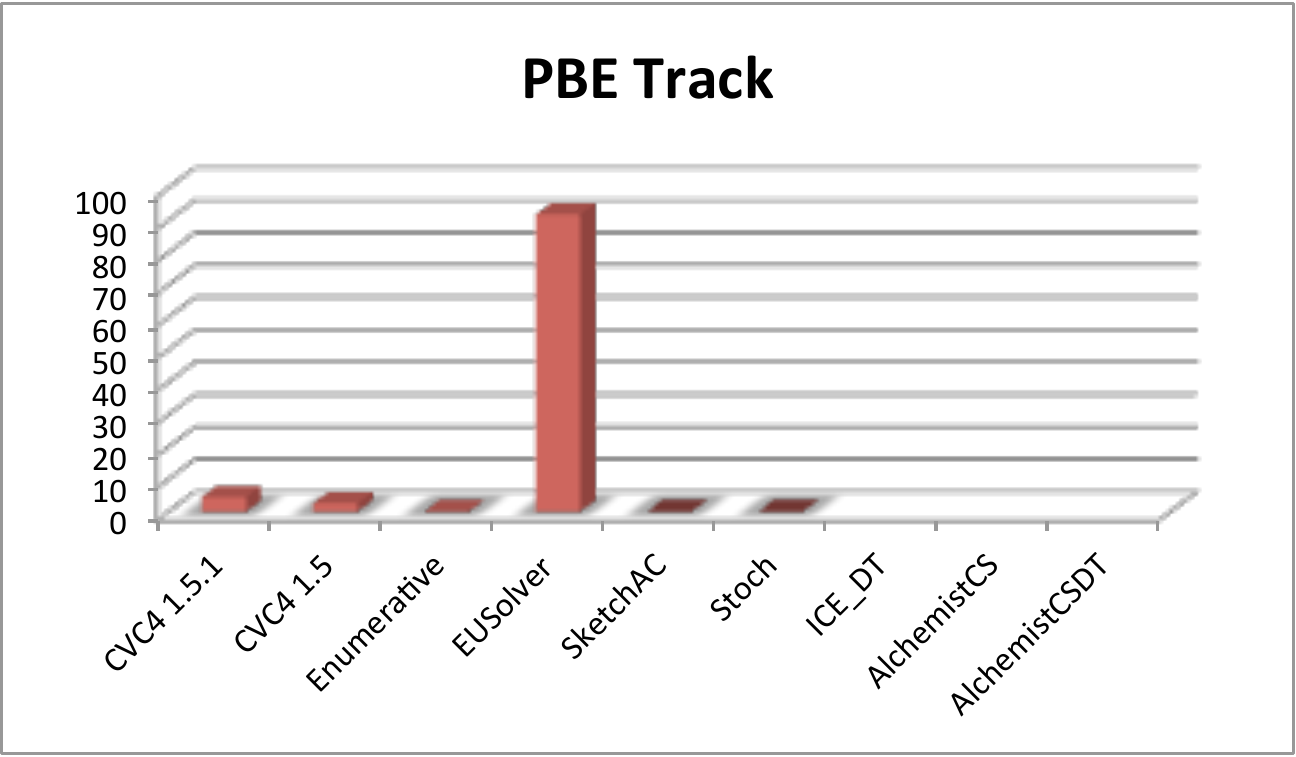}
		\label{fig:pbeoverall}
	\end{minipage}
	\end{center}
	\caption{The percentage of benchmarks solved by the different solvers in each of the four tracks}
	\label{fig:percentagePerTrack}	
\end{figure}

\begin{figure*}
		\noindent\makebox[\textwidth]{
			\scalebox{0.6}{
				\begin{tabular}{c}
					\includegraphics[width=9.5in,bb=7 9 938 469]{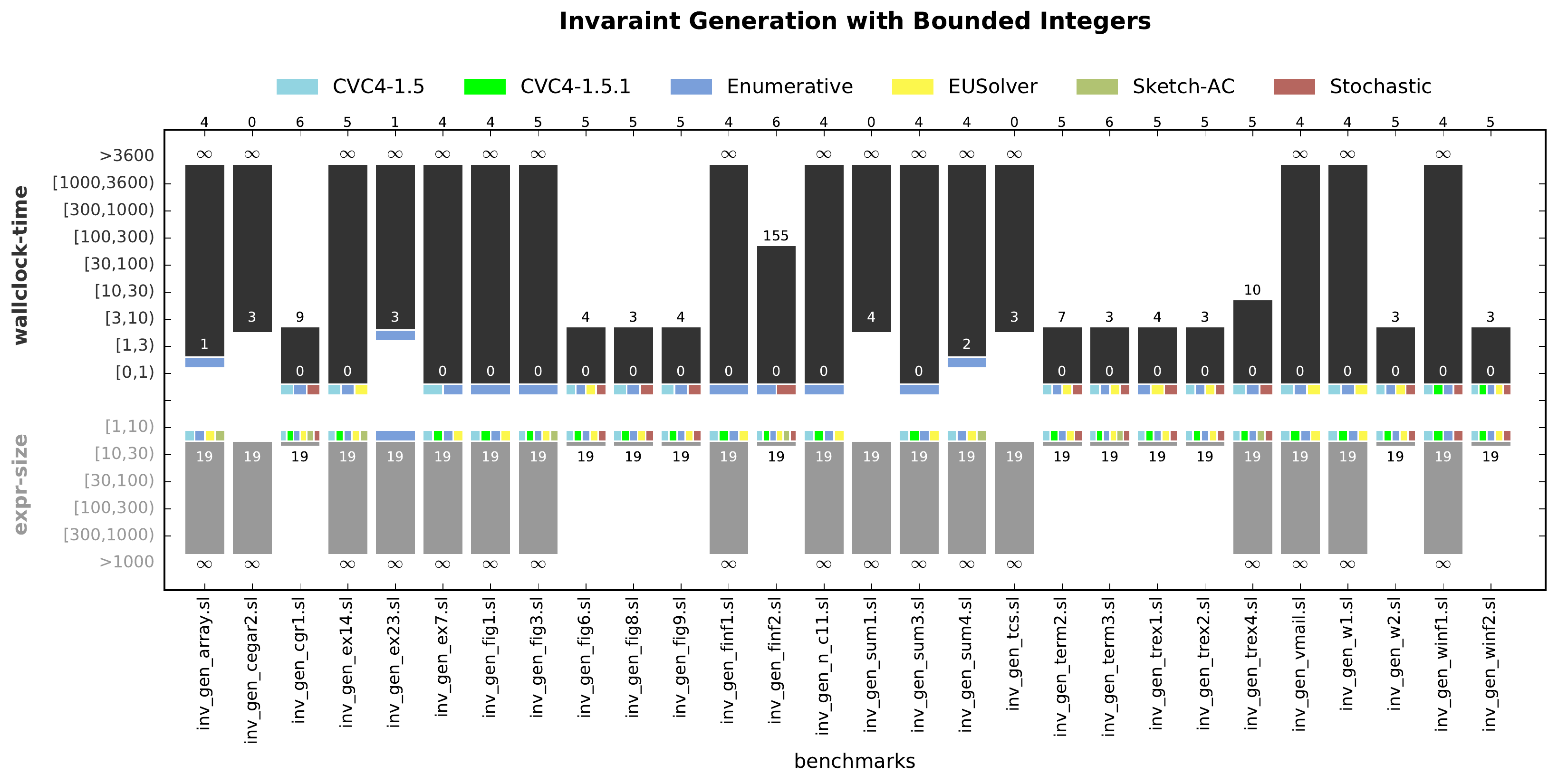} \\
					\includegraphics[width=9.5in,bb=7 9 938 505]{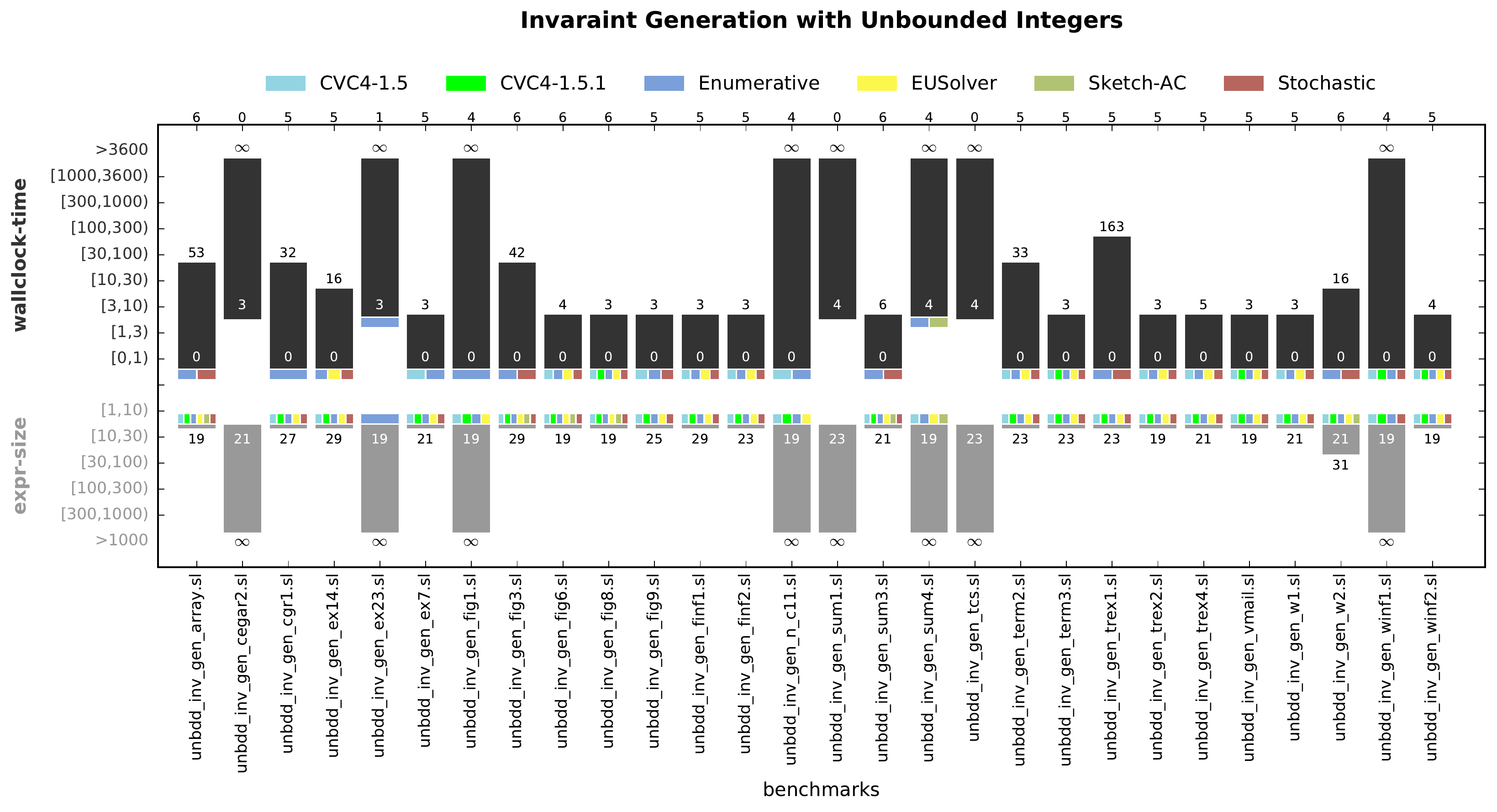} 
				\end{tabular}
			}}
			\caption{Evaluation of Invariant benchmarks of the general track.}\label{fig:inv-gnrl-results}
		\end{figure*}

\begin{figure*}
		\noindent\makebox[\textwidth]{
			\scalebox{0.6}{
				\begin{tabular}{c}
					\includegraphics[width=9.5in,bb=7 9 923 469]{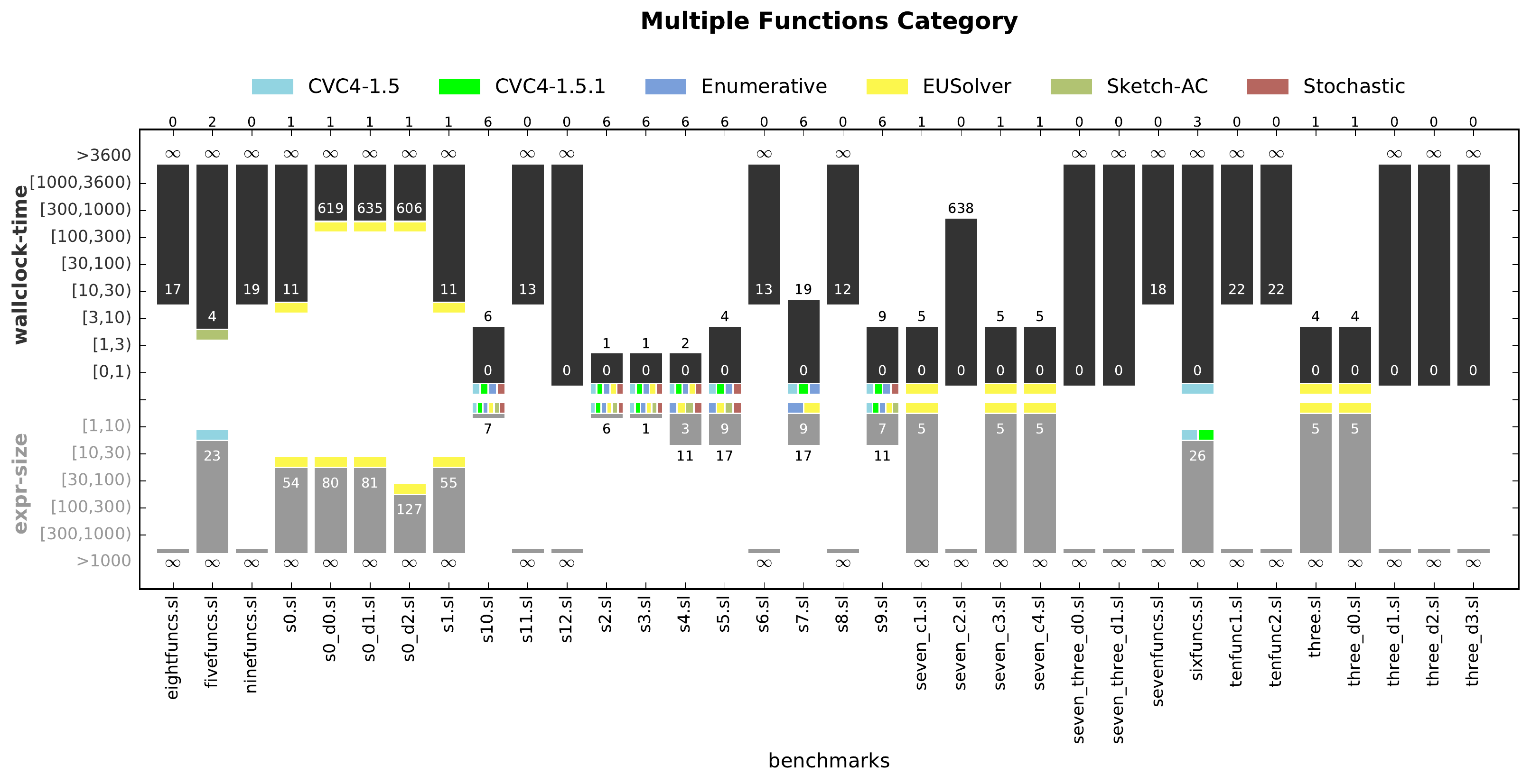} \\
					\includegraphics[width=9.5in,bb=7 9 930 478]{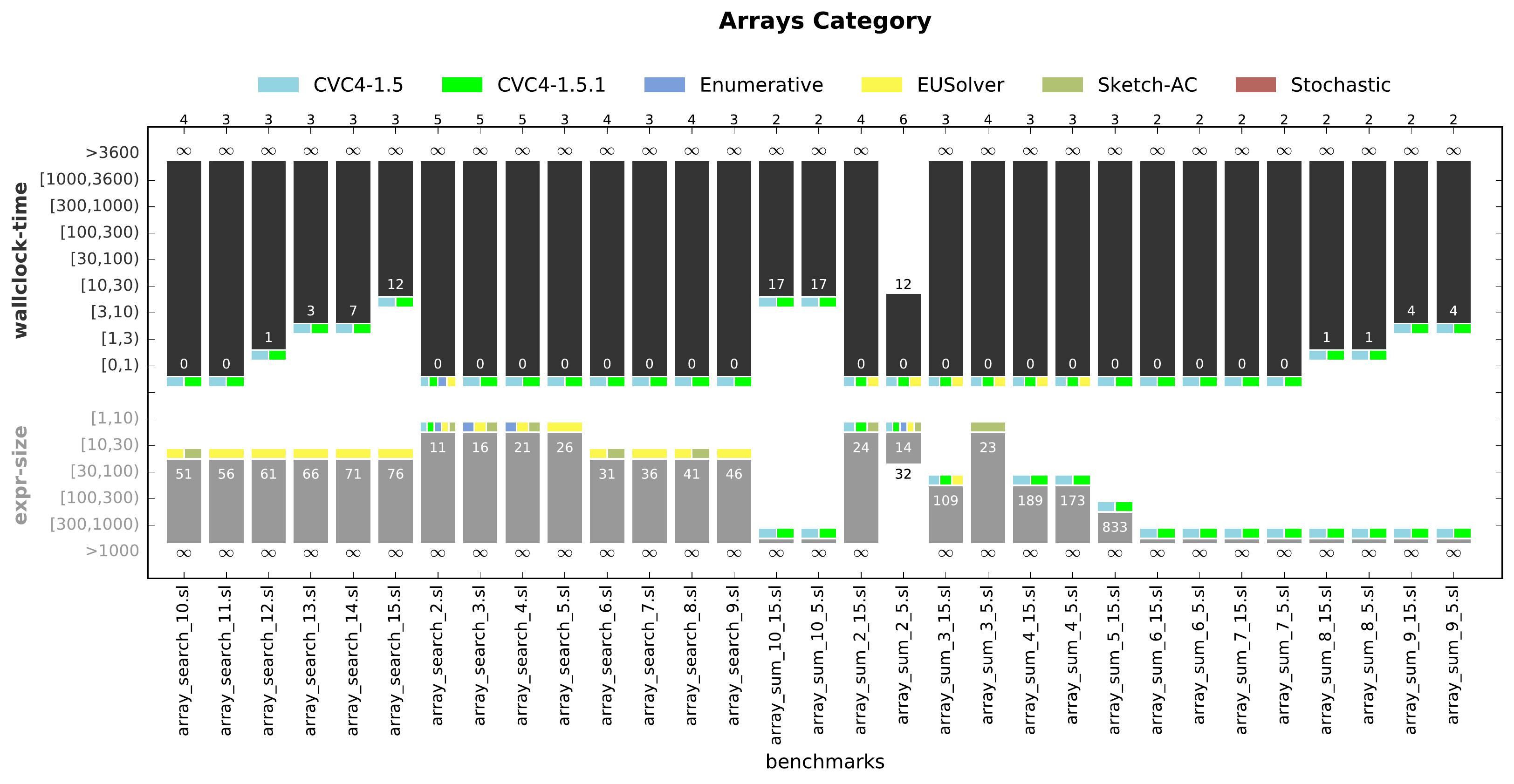} 
				\end{tabular}
			}}
			\caption{Evaluation of Multiple Functions and Arrays benchmarks.}\label{fig:mult-arrays-results}
		\end{figure*}

\begin{figure*}
			\noindent\makebox[\textwidth]{
				\scalebox{0.6}{
					\begin{tabular}{c}
						\includegraphics[width=9.5in,bb=8 9 911 462]{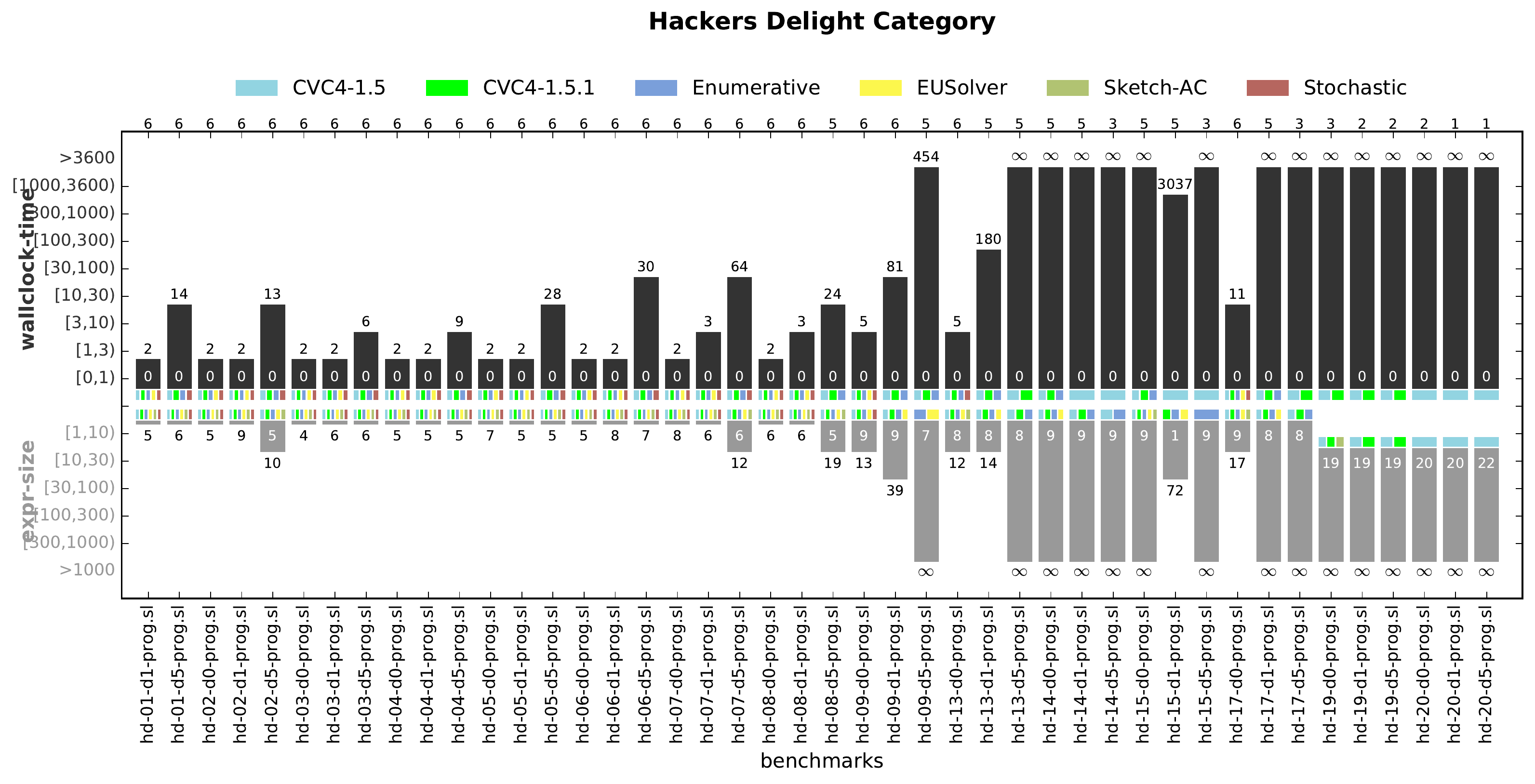} \\
						\includegraphics[width=9.5in,bb=7 9 923 471]{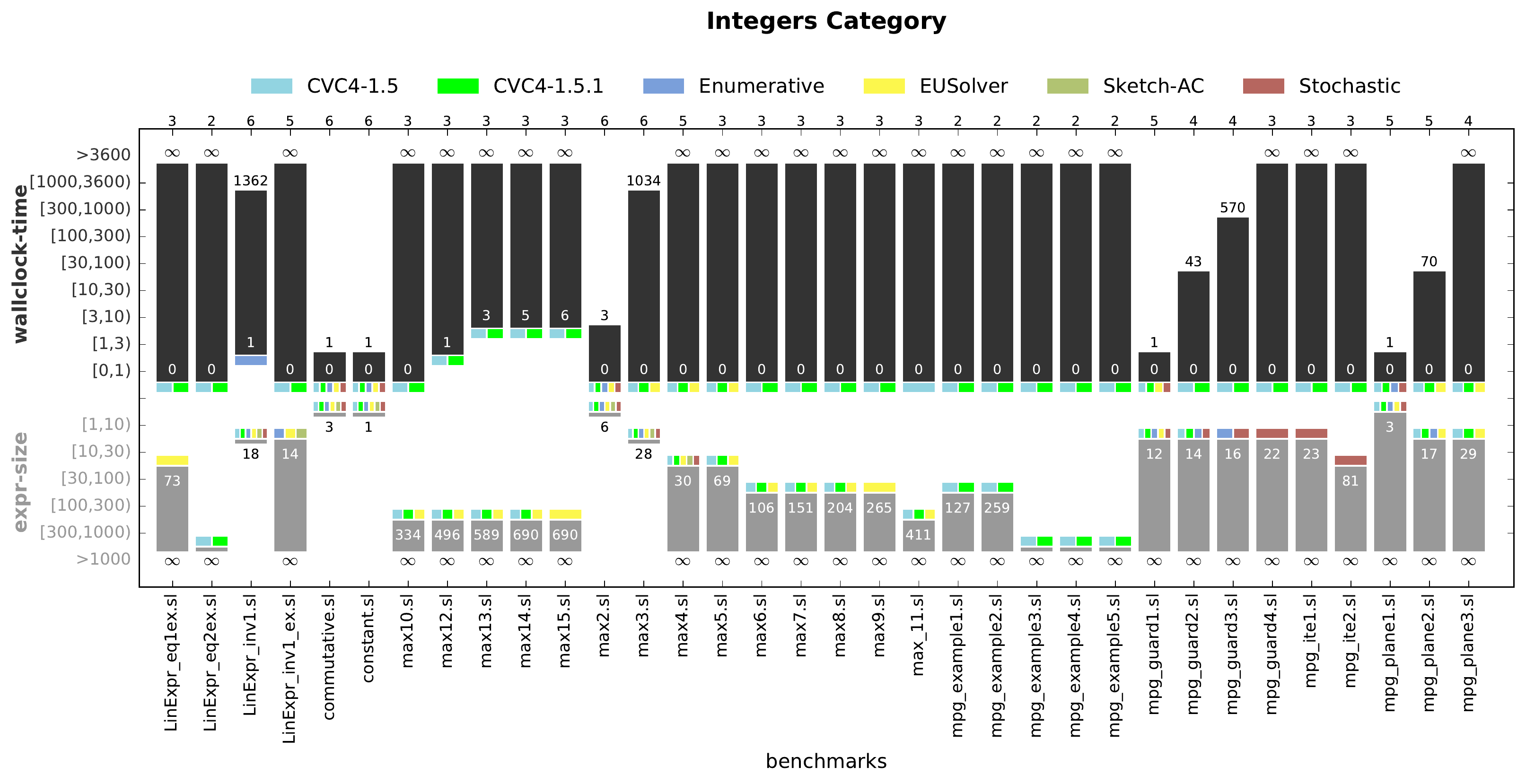} 
					\end{tabular}
				}}
				\caption{Evaluation of Hackers Delight and Integer benchmarks.}\label{fig:hd-int-results}
			\end{figure*}

The percentage of benchmarks solved by each solver in the General track is shown in Figure~\ref{fig:percentagePerTrack} on the top left. The $\eusolver$ solves the maximum number of benchmarks 206 out of 309. The $\cvcnew$ solver solves 195 benchmarks, whereas the last year's winner in this category $\cvc$ solved 179 benchmarks. The $\eusolver$ solved 59 benchmarks uniquely and $\cvcnew$ solved 22 benchmarks uniquely. With regard to time to solve, the $\cvcnew$ solvers solved 161 benchmarks among the fastest whereas the $\eusolver$ solved 127 benchmarks among the fastest. For details on the expression size see Figures~\ref{fig:co-bv-let-mp-results} to~\ref{fig:hd-int-results}.

\paragraph{Conditional Linear ArithmeticTrack}

The percentage of benchmarks solved by the solvers in the Conditional Linear Integer Arithmetic track is shown in Figure~\ref{fig:percentagePerTrack} on the top right. The $\cvcnew$ solver solved all 73 benchmarks in this category, whereas the $\eusolver$ solved 72 out of the 73 benchmarks. Last year's winner in this category, $\cvc$, solved 70 benchmarks. One benchmark was solved uniquely, by $\cvcnew$. The $\cvcnew$ solver solved 72 benchmarks among the fastest and $\eusolver$ solved 33 among the fastest.

\paragraph{Invariant Synthesis Track}
The result for the invariant synthesis track is shown in Figure~\ref{fig:percentagePerTrack} on the bottom left. In this track, the $\ice$ solver (also last year's track winner) solves the maximum number of benchmarks 57 out of 67. The $\cvcnew$ solver solves 56 benchmarks, whereas the $\alccsdt$ solver solves 52 benchmarks. Two benchmarks were solved uniquely, the two by $\ice$. In terms of time to solve $\cvcnew$ preformed best, solving 50 bechmarks among the fastest. This is an impressive improvement from last years' version $\cvc$ which solved 10 benchmarks among the fastest. The $\ice$ solver solved 44 benchmarks among the fastest and the $\alccsdt$ solver solved 37 benchmarks among the fastest.

\paragraph{Programming By Example Track}

The results for the new Programming By Example (PBE) track is shown in Figure~\ref{fig:percentagePerTrack} on the bottom right. Unlike other tracks, we see a dramatic difference in the performance of the solvers for the benchmarks in the PBE track. The $\eusolver$ remarkably solves 787 benchmarks out of 858 (742 out of 745 in the bit-vectors category and 45 out of 108 in the strings category), whereas the second best solver $\cvcnew$ solves 39 benchmarks (21 in the bit-vectors category and 18 in the strings category). No other solver solved more than 1 problem in this track. The $\eusolver$ solved 751 benchmarks uniquely (720 in the bit-vectors category and 31 in the strings category), and $\cvcnew$ solved 4 benchmarks uniquely (all in the strings category).

\subsection{Detailed Results}
\label{subsec:benchs-pres}
In the following section we show the results of the competition from the benchmark's perspective. 
For a given benchmark we would like to know: how many solvers solved it, what is the min and max time to solve,  what are the min and max size of the expressions produced, which solver solved the benchmark the fastest, and which solver produced the smallest expression.

	\begin{figure*}
		\noindent\makebox[\textwidth]{
			\scalebox{0.6}{
				\begin{tabular}{c}
					\includegraphics[width=9.5in,bb=7 9 944 469]{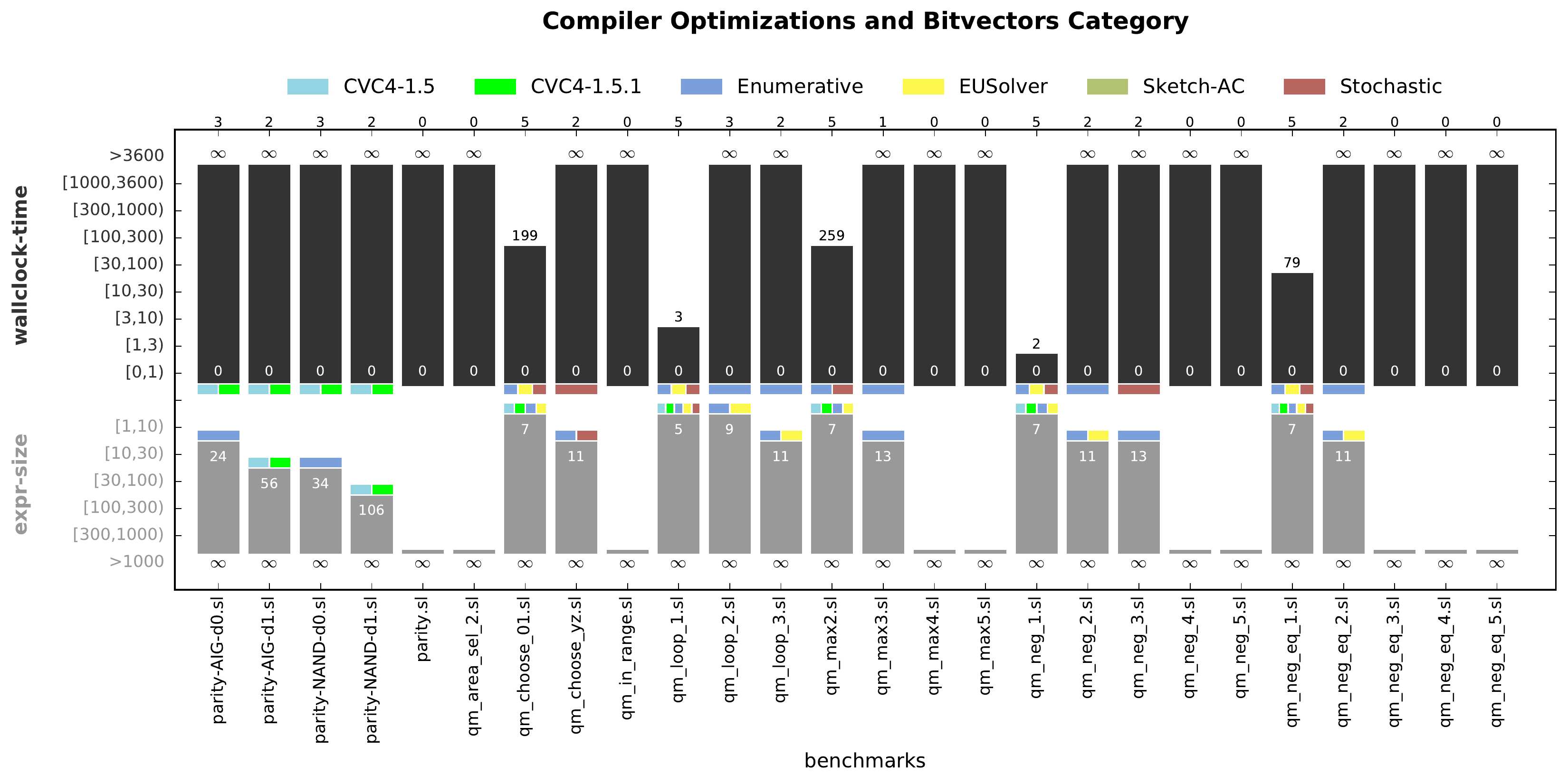} \\
					\includegraphics[width=9.5in,bb=7 9 941 456]{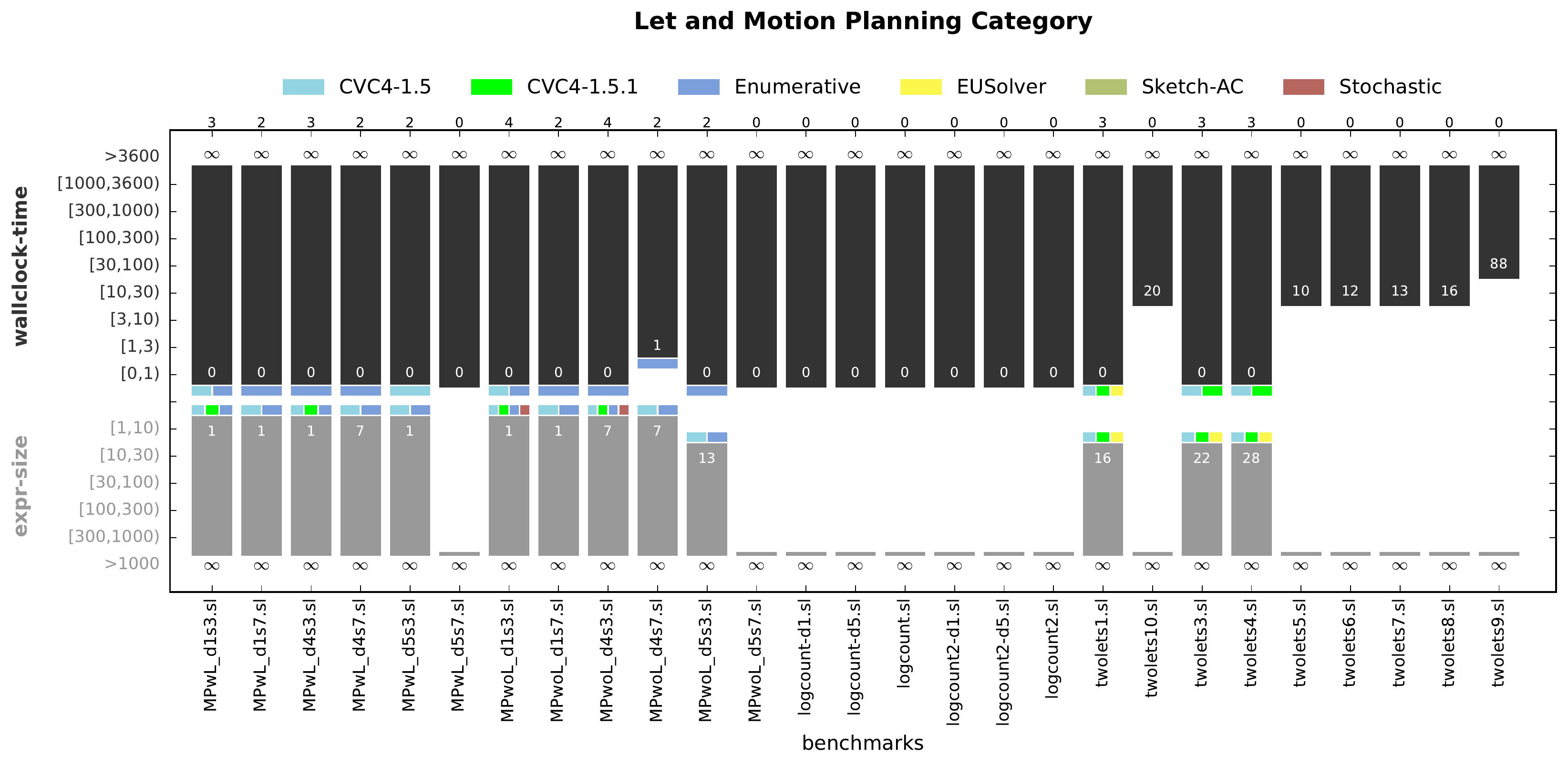} 
				\end{tabular}
			}}
			\caption{Evaluation of Compiler Optimizations, BitVectors, Lets and Motion Planning benchmarks.}\label{fig:co-bv-let-mp-results}
		\end{figure*}

	We represents the results per benchmarks in groups organized per tracks and categories. For instance, Fig.~\ref{fig:co-bv-let-mp-results} at the top presents details of the compiler optimization benchmarks. The black bars show the range of the time to solve among the different solvers in pseudo logarithmic scale (as indicated on the upper part of the y-axis). Inspect for instance benchmark \texttt{qm\_choose\_01.sl}. The black bar indicates that the fastest solver to solve it used less than 1 second, and the slowest used between 100 to 300 seconds. 
	The black number above the black bar indicates the exact number of seconds (floor-rounded to the nearest second) it took the slowest solver to solve a benchmark (and $\infty$ if at least one solver exceeded the time bound). Thus, we can see that the slowest solver to solve \texttt{qm\_choose\_01.sl} took 199 seconds to solve it. The white number at the lower part of the bar indicates the time of the fastest solver to solve that benchmark. Thus, we can see that the fastest solver to solve \texttt{qm\_choose\_01.sl} required less than 1 second to do so. The colored squares/rectangles next to the lower part of the black bar, indicate which solvers were the fastest to solve that benchmark (according to the solvers' legend at the top). Here, \emph{fastest} means in the same logarithmic scale as the absolute fastest solver. For instance, we can see that \enum, \stoch, and \eusolver\ were the fastest to solve \texttt{qm\_choose\_01.sl}, solving it in less than a second
	and that among the 2 solvers that solved \texttt{MPwoL\_d5s3.sl} only \enum\ solved it in less than 3 seconds. 
	
	Similarly, the gray bars indicate the range of expression sizes in pseudo logarithmic scales (as indicated on the lower part of the y-axis), where the size of an expression is determined by the number of nodes in its parse tree.
	The black number at the bottom of the gray bar indicates the exact size expression of the largest solution (or $\infty$ if it exceeded 1000), and the white number at the top of the gray bar indicates the exact size expression of the smallest solution. The colored squares/rectangles next to the upper part of the gray bar indicates which solvers (according to the legend) produced the smallest expression (where \emph{smallest} means in the same logarithmic scale as the absolute smallest expression). For instance, for \texttt{qm\_choose\_01.sl} the smallest expression produced had size 7, and 4 solvers out of the 5 who solved it managed to produce an expression of size less than 10.  
	
	Finally, at the top of the figure above each benchmark there is a number indicating the number of solvers that solved that benchmark. For instance, one solver solved \texttt{qm\_max3.sl}, two solvers solved \texttt{qm\_neg2.sl}, three solvers solved \texttt{qm\_loop2.sl}, and no solver solved \texttt{qm\_max4.sl} or \texttt{twolets10.sl}. Note that the reason \texttt{twolest10.sl} has 20 as the lower time bound, is that that is the time to terminate rather than the time to solve. Thus, one of the solvers has terminated within 20 seconds, but either it did not produce a result, or it produced an incorrect result. When no solver produced a correct result, there will be no colored squares/rectangles next to the lower parts of the bars.

	\begin{figure*}
		\noindent\makebox[\textwidth]{
			\scalebox{0.6}{
				\begin{tabular}{c}
					\includegraphics[width=9.5in,bb=7 9 917 493]{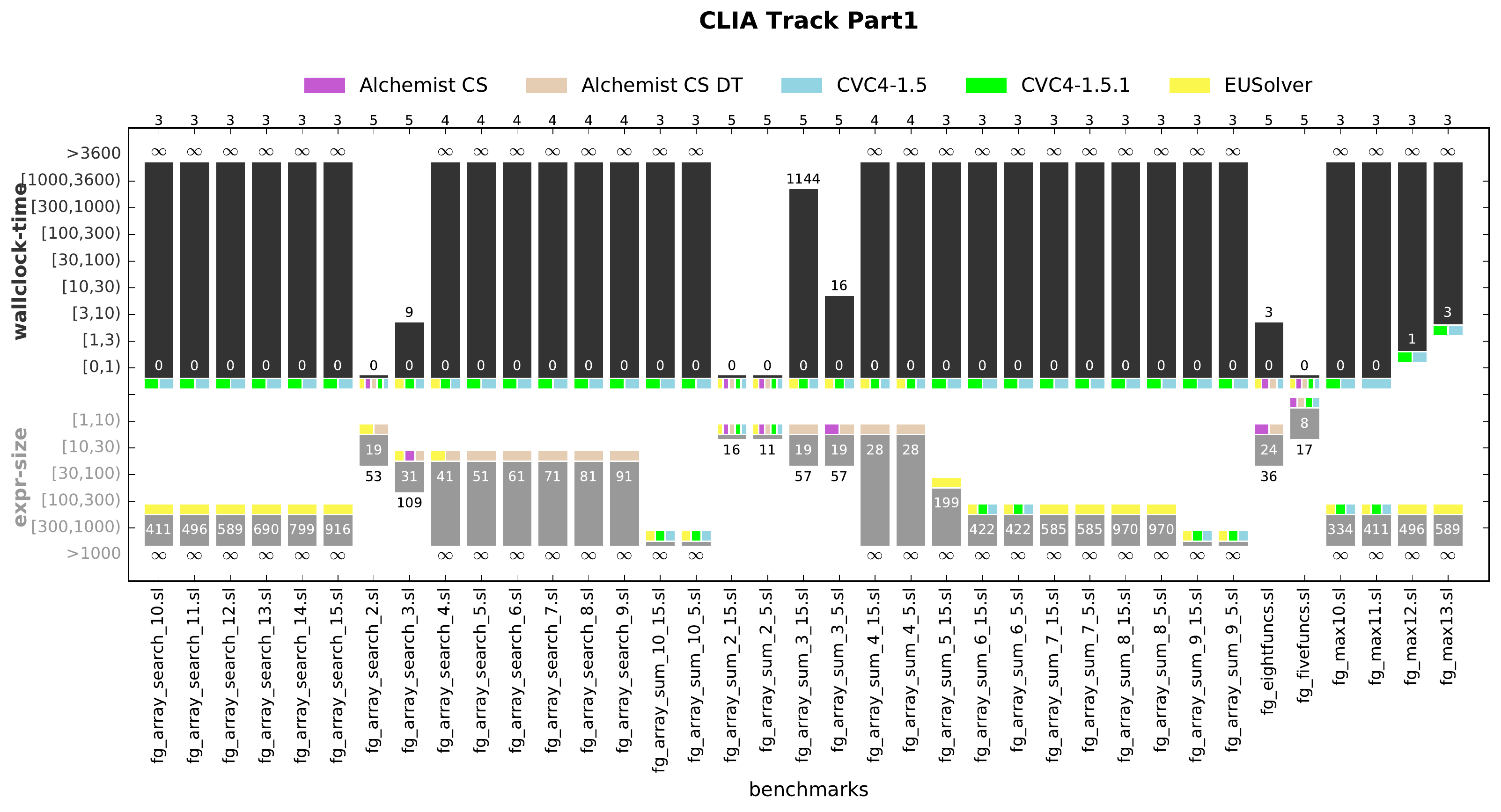} \\
					\includegraphics[width=9.5in,bb=7 9 919 482]{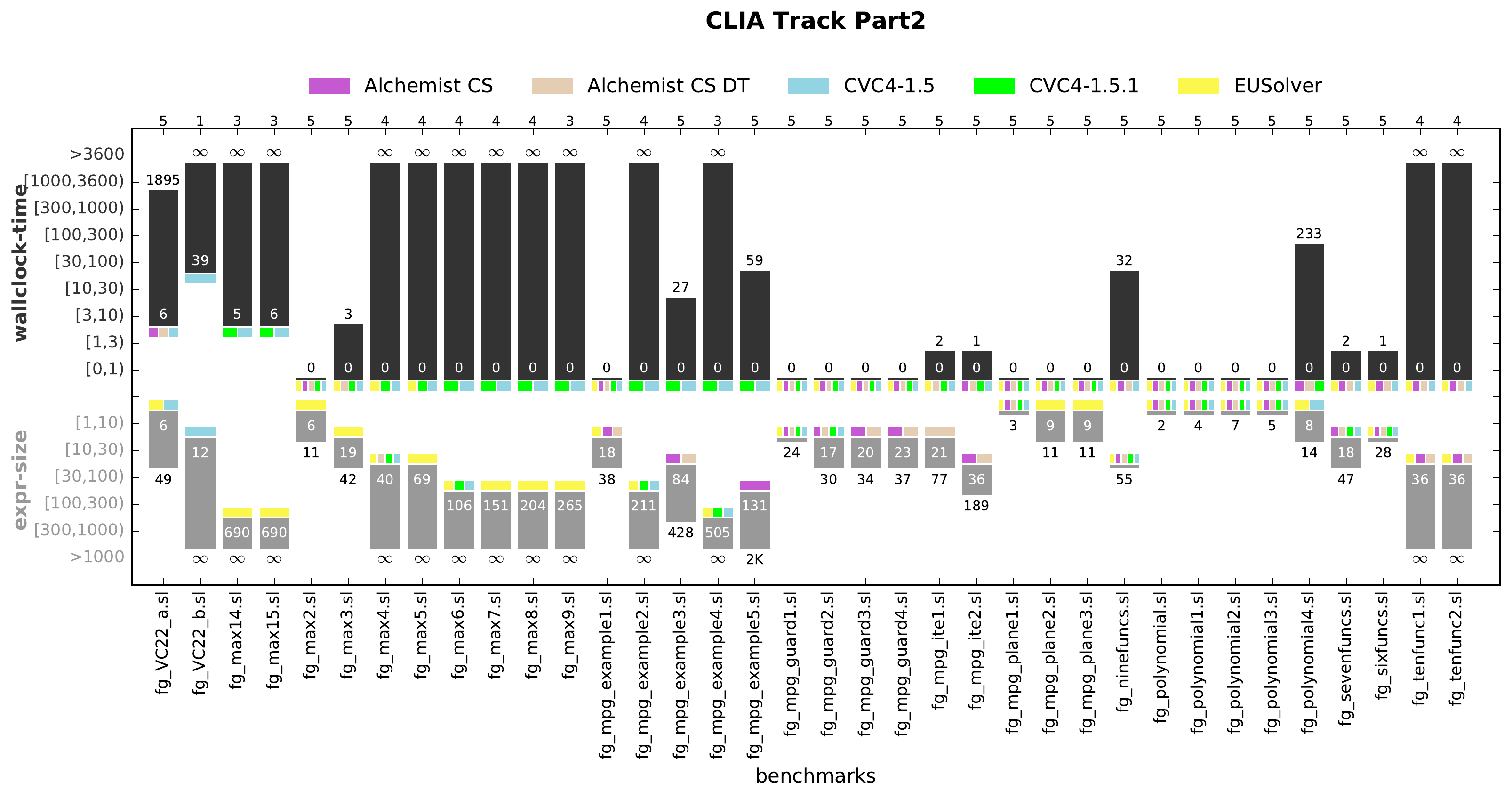} 
				\end{tabular}
			}}
			\caption{Evaluation of CLIA track benchmarks.}\label{fig:lia-results}
		\end{figure*}

	\begin{figure*}
		\noindent\makebox[\textwidth]{
			\scalebox{0.6}{
				\begin{tabular}{c}
					\includegraphics[width=9.5in,bb=7 9 923 476]{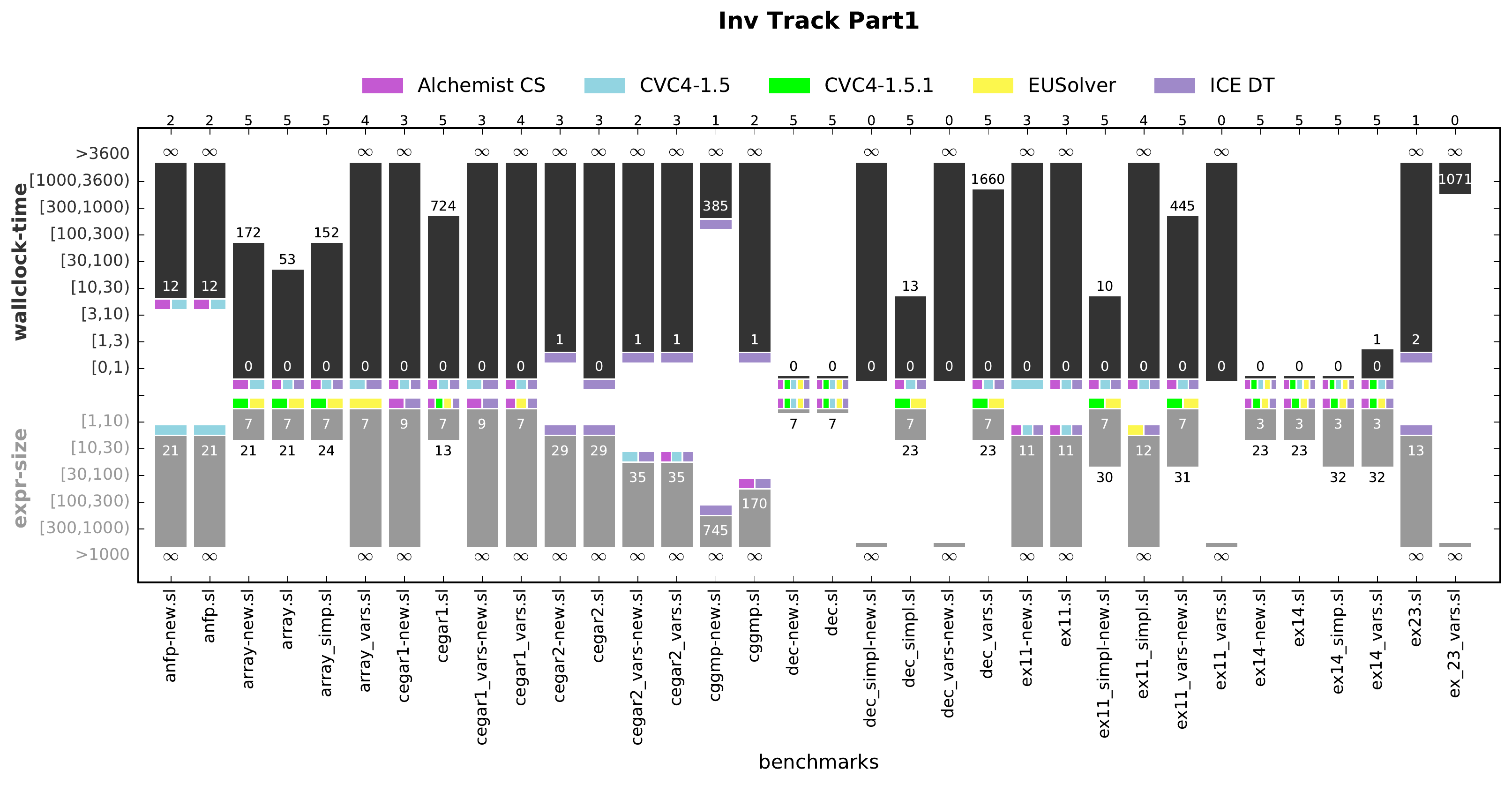} \\
					\includegraphics[width=9.5in,bb=7 9 925 460]{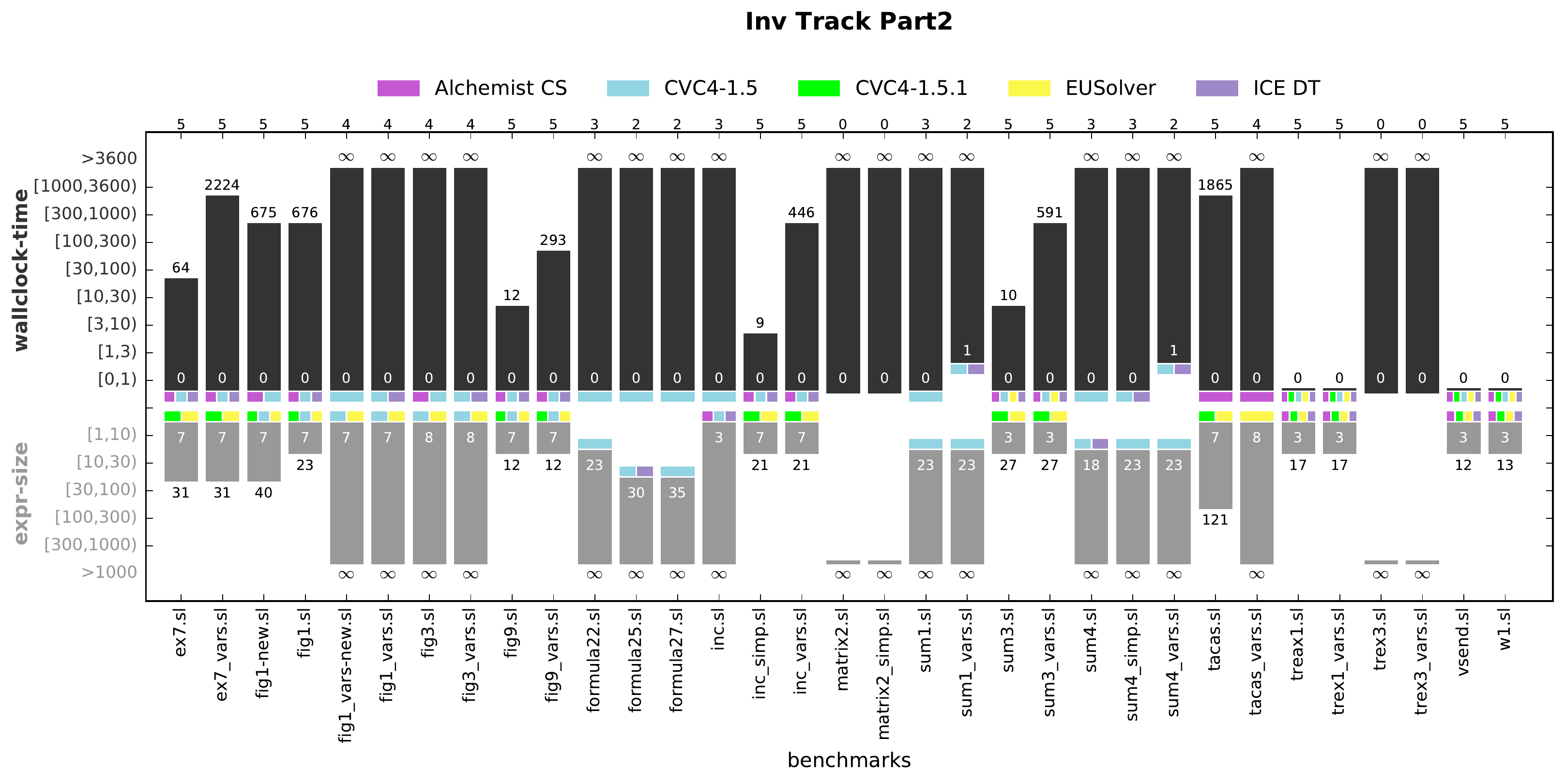} 
				\end{tabular}
			}}
			\caption{Evaluation of Invariant track benchmarks.}\label{fig:inv-results}
		\end{figure*}

	\begin{figure*}
		\noindent\makebox[\textwidth]{
			\scalebox{0.6}{
				\begin{tabular}{c}
					\includegraphics[width=9.5in,bb=7 9 913 537]{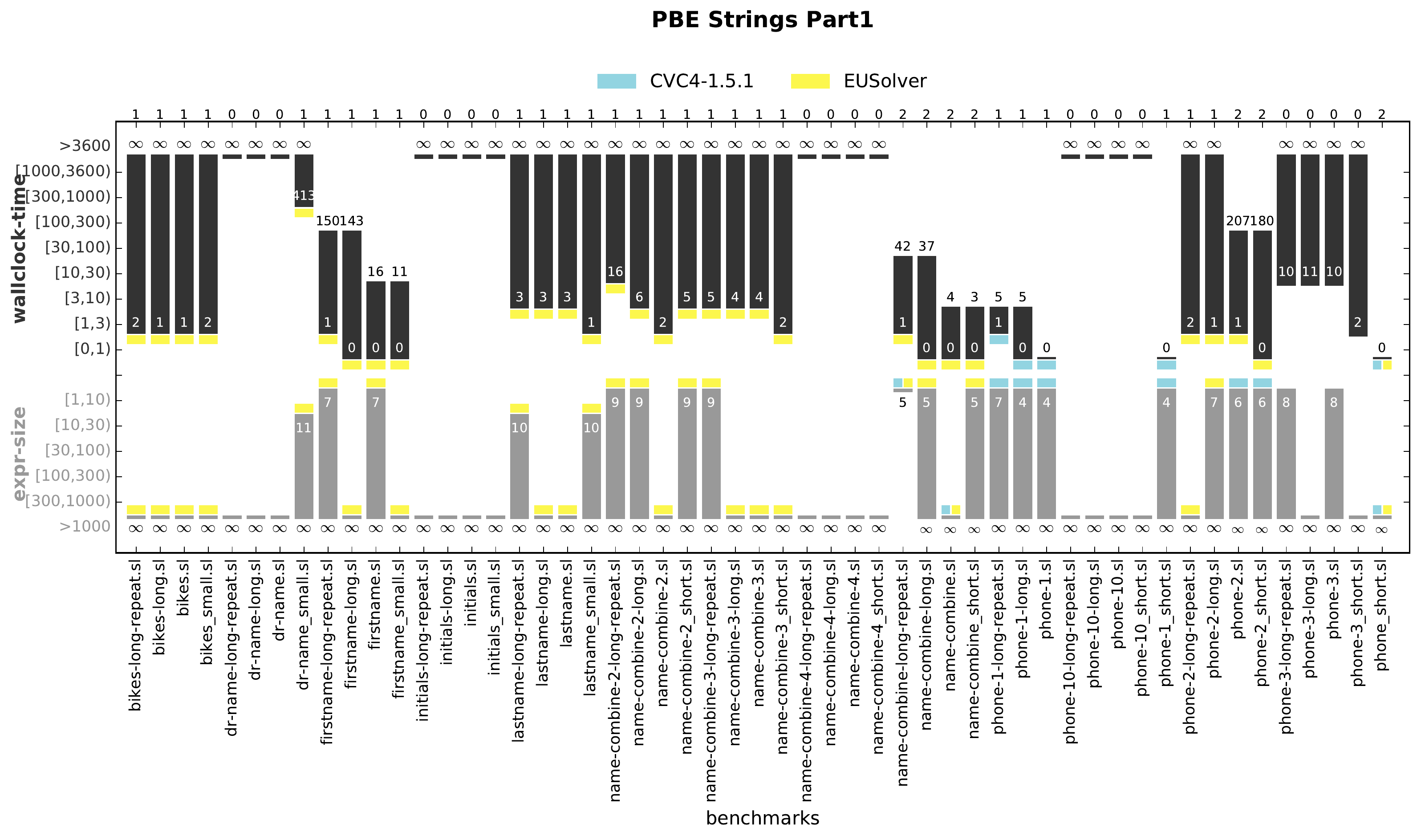} \\
					\includegraphics[width=9.5in,bb=7 9 911 522]{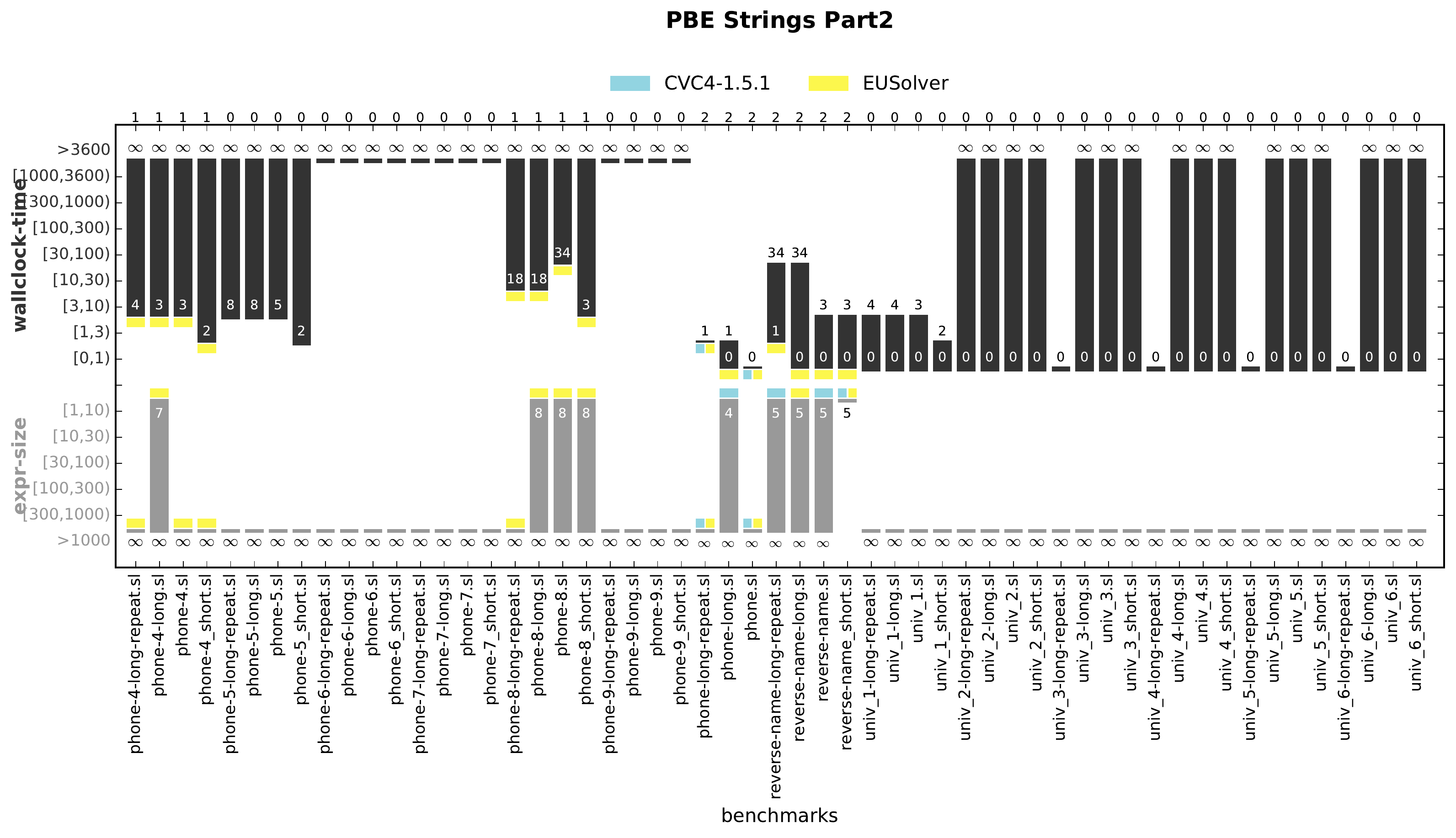} 
				\end{tabular}
			}}
			\caption{Evaluation of PBE-Strings benchmarks.}\label{fig:pbe-strings-results}
		\end{figure*}

\subsection{Observations} 
	\begin{figure}[t]
	\begin{center}
		\begin{minipage}{.485\textwidth}
			\centering
			\includegraphics[scale=0.325,bb=0 0 670 429]{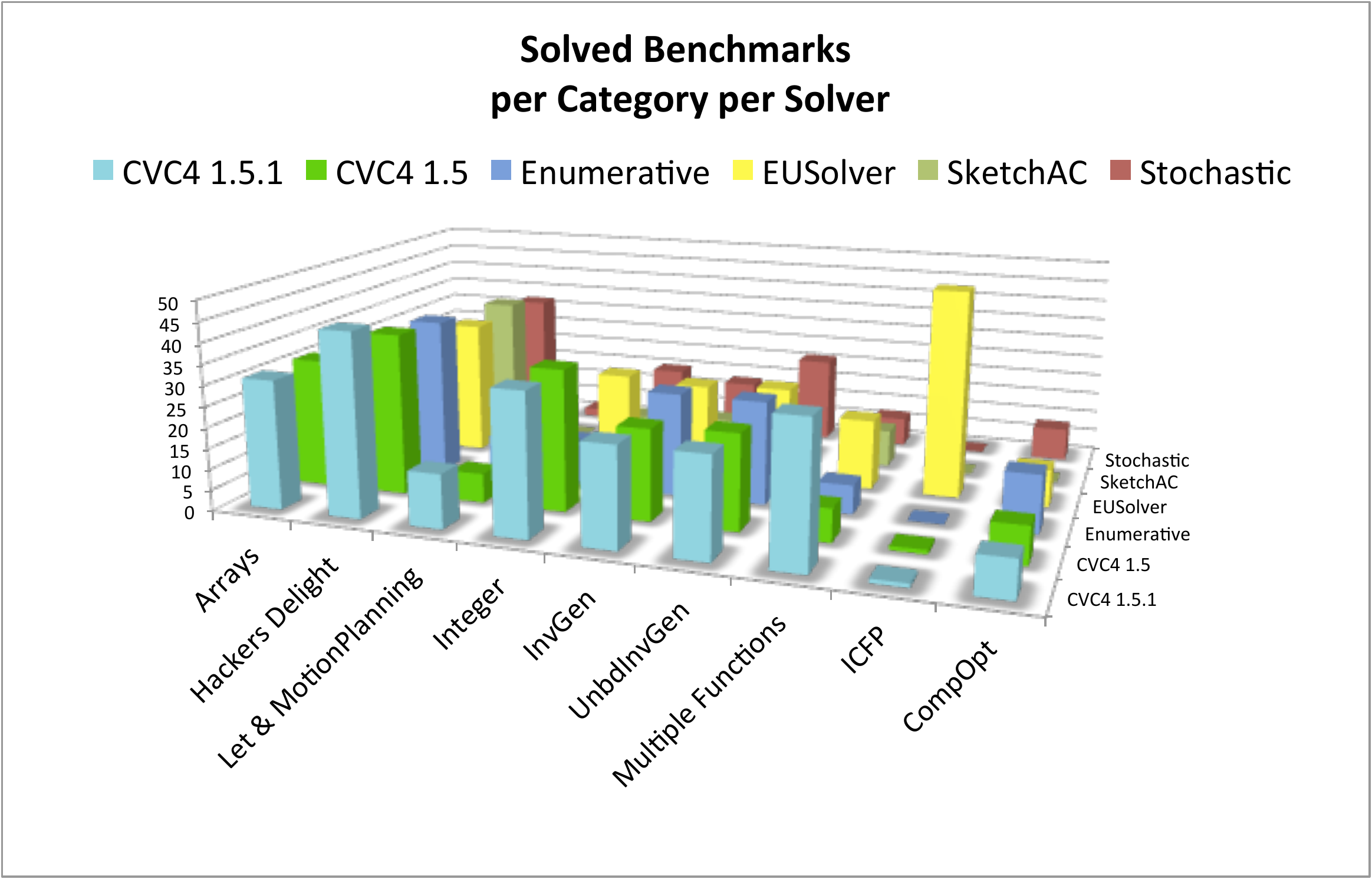}
		\end{minipage}
		\begin{minipage}{.485\textwidth}
			\centering
			\includegraphics[scale=0.325,bb=0 0 668 429]{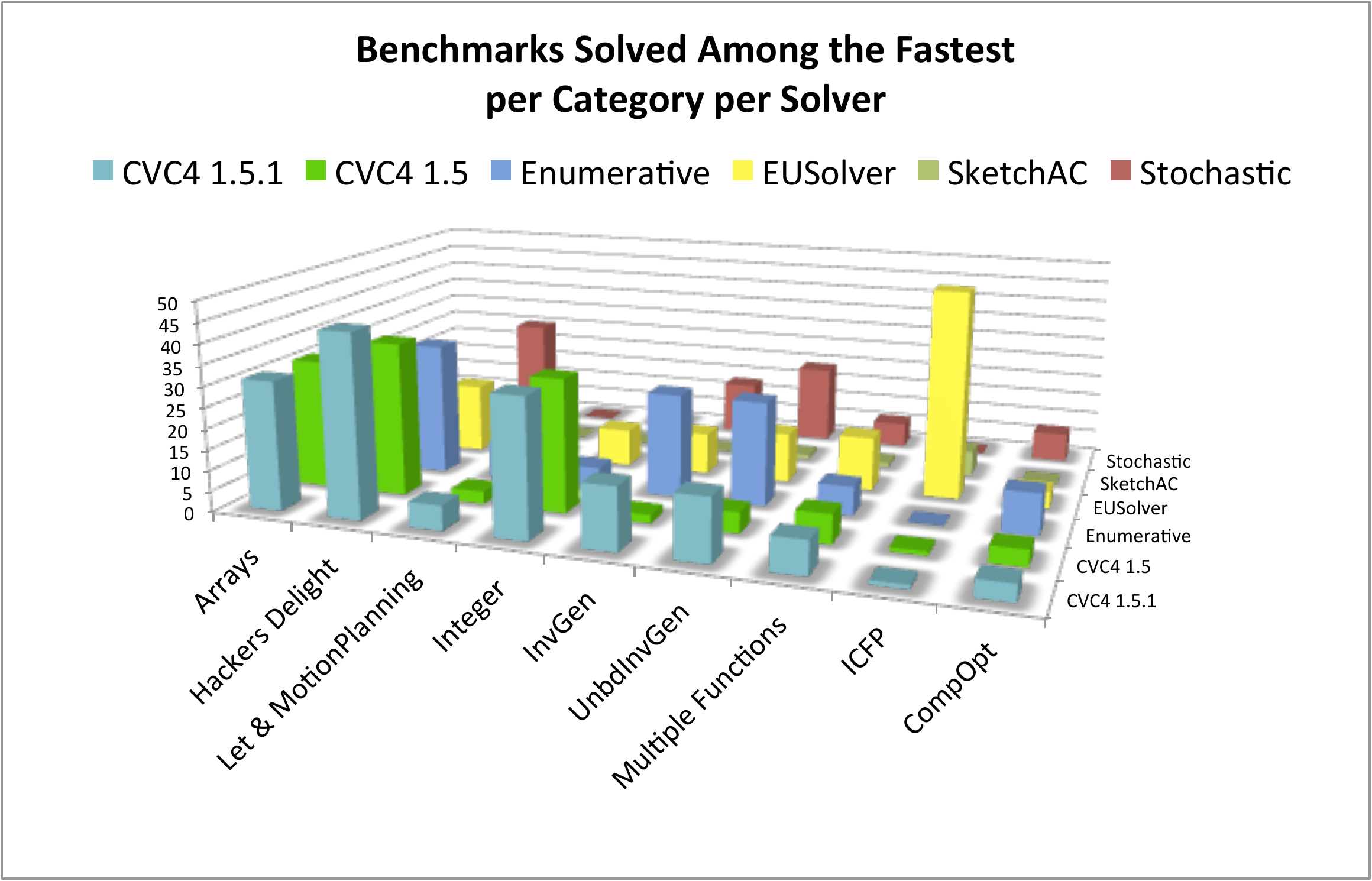}
		\end{minipage}
	\end{center}
	\caption{Results of General Tracks per Solver per Category.}
	\label{fig:general-per-cat}	
	\end{figure}
Analyzing the results of the general track per category (see Figure~\ref{fig:general-per-cat}), along the number of benchmarks solved, the number of benchmarks solved uniquely and the number of benchmarks solved among the fastest, we can see that each category of the general track has a clear winner:
\begin{itemize}
	\item $\cvcnew$ won 4 categories: Arrays, Let \& Motion Planning, Hackers' Delight and Integers.
	\item $\enum$ won 3 categories: Compiler-Optimizations, Invariant Generation and Invariant Generation with unbounded integers.
	\item $\eusolver$ won 3 categories: Multiple Functions and ICFP.
\end{itemize}
If we disregard the partition to categories we can make the following observations:
\begin{itemize}
	\item \eusolver\ solved more benchmarks of the general track than all other solvers
		\begin{itemize} 
			\item $\eusolver$ solved 206/309,
			\item $\cvcnew$ solved 195/306, and
			\item $\enum$ solved 139/309.
		\end{itemize}
	\item In terms of time to solve, $\cvcnew$ solved more benchmarks among the fastest
		\begin{itemize} 
			\item $\cvcnew$ solved 157 among fastest,
			\item $\eusolver$ solved 123 among fastest, and
			\item $\enum$ solved 114 among fastest.
		\end{itemize}	
\end{itemize}

With regard to expression sizes, we see that in average $\cvcnew$ and $\eusolver$ generate large expressions. The average expression size for $\cvcnew$ is 31580.5 and for $\eusolver$ it is 30595.7 whereas the average sizes of $\enum$, $\skac$ and $\stoch$ are between 11.9 to 17.1. This comparison is not particularly fair, since both $\cvcnew$ and $\eusolver$ solved more benchmarks in general, so that might be the reason. For this reason we give the exact size expression per benchmark in the detailed evaluation figures (Figs.~\ref{fig:co-bv-let-mp-results} to~\ref{fig:hd-int-results}). Looking at these figures we can see that in many instances where the benchmark was solved by both $\cvcnew$ and $\eusolver$, the size of the expression generated by $\eusolver$ was in a smaller bucket according to our pseudo-logarithmic scale, see for instance the \texttt{array\_search*} benchmarks and the \texttt{fg\_max*} benchmarks.

\section{Discussion}
\label{sec:discussion}
\begin{figure}
\centering
\includegraphics[trim=3cm 18.6cm 0cm 0cm,bb=96 505 532 754]{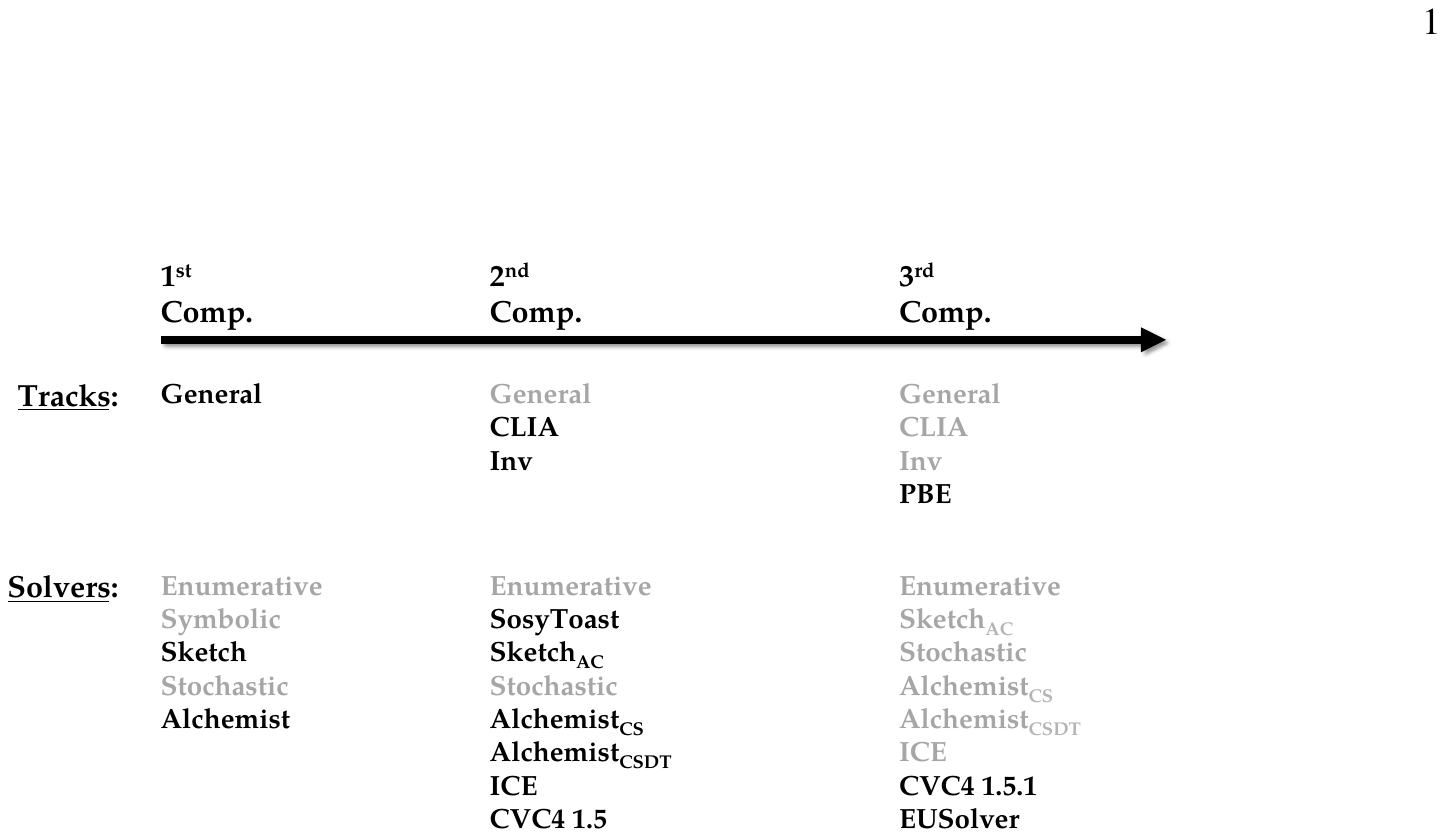}
\caption{The timeline for tracks and solvers for each competition.}
\label{timelinesolvers}
\end{figure}

We  present a few interesting dimensions in which the SyGuS competition has evolved over the past 3 years. The timeline for the tracks and the solvers submitted for each competition is shown in Figure~\ref{timelinesolvers}. The first competition in 2014 had a single General track, and 5 solvers competed in the competition that included enumerative, stochastic, symbolic, and machine learning-based synthesis algorithms. The second competition introduced two new tracks: conditional linear integer arithmetic track and the invariant synthesis track. There were 7 new solver submissions that implemented SMT-based quantifier instantiation, adaptive concretization of unknowns, BDD-based symbolic algorithms, and geometric optimization based synthesis algorithms. In the 2016 competition, we introduced another new track, the PBE track, and two new solvers $\eusolver$ and $\cvcnew$ participated in the competition.

\begin{figure}
		\begin{minipage}{.45\textwidth}
			\centering
			\includegraphics[scale=0.4,bb=0 0 491 319]{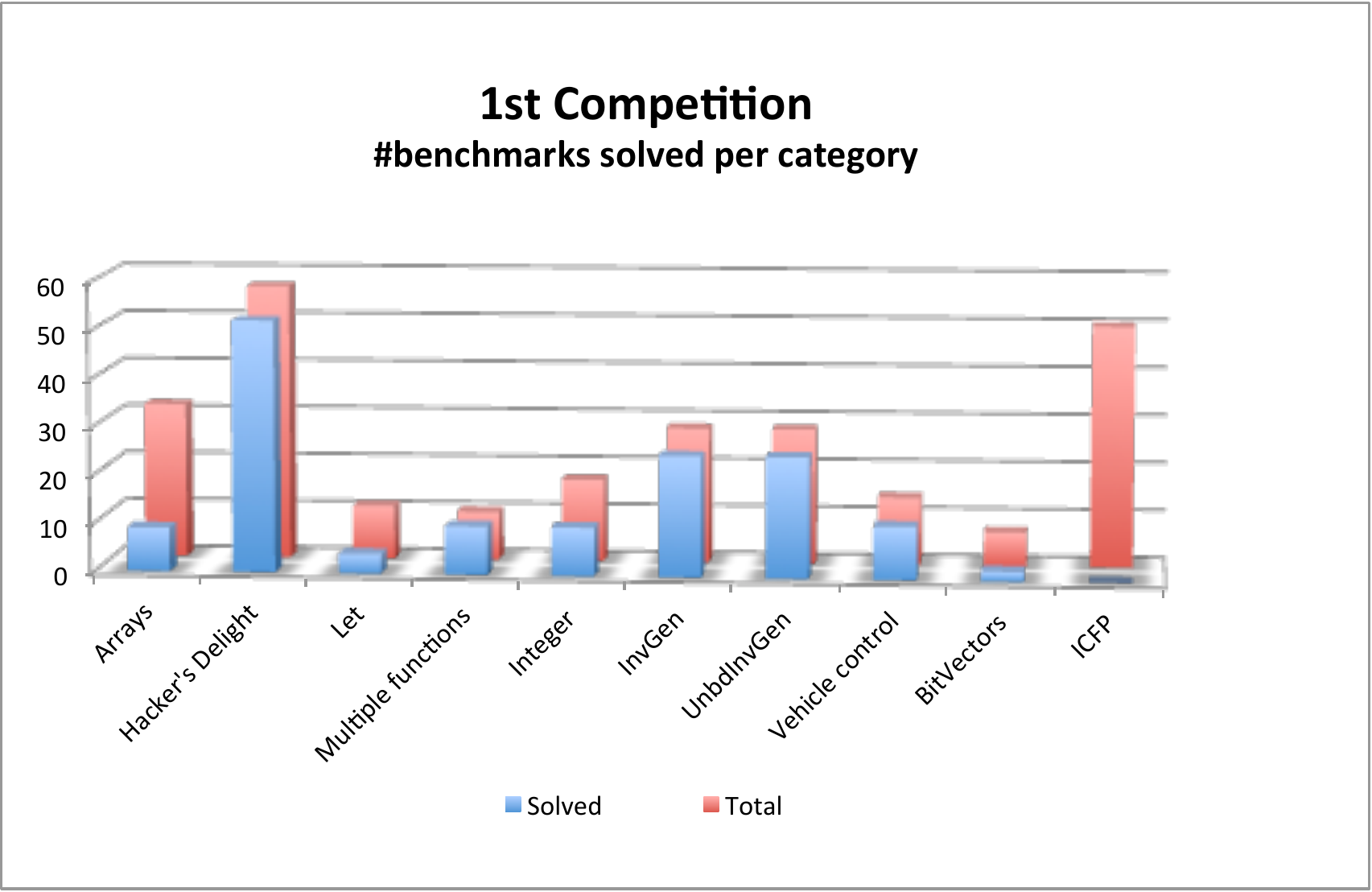}
			\label{fig:generaloverall}
		\end{minipage}
		\begin{minipage}{.45\textwidth}
			\centering
			\includegraphics[scale=0.4,bb=0 0 491 319]{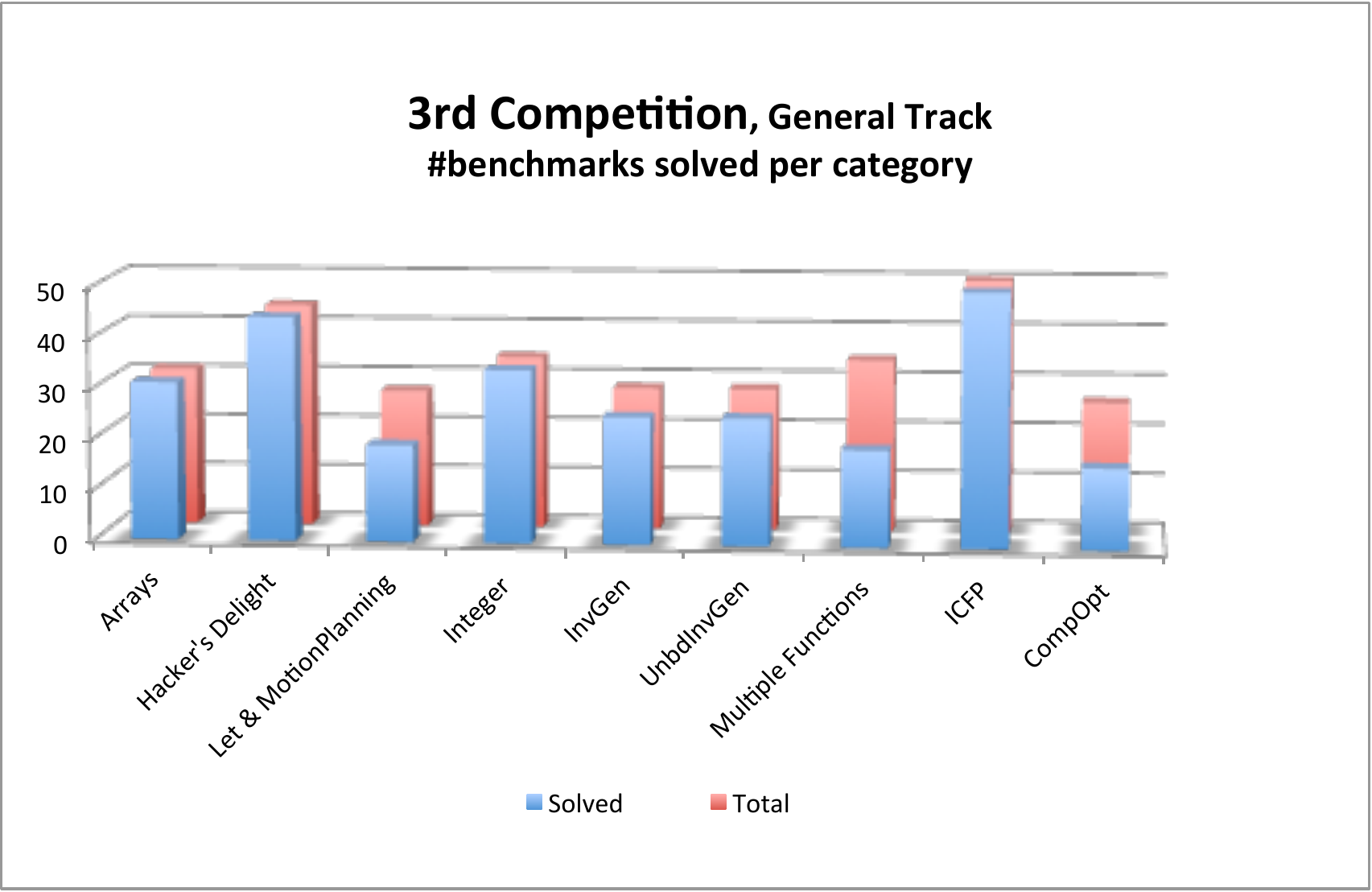}
			\label{fig:liaoverall}
		\end{minipage}
	\caption{The comparison between the number of benchmarks solved out of the total in the 1st competition and the general track of the 3rd competition.}
	\label{benchmarksfirstthird}
\end{figure}

\begin{figure}
\centering
\includegraphics[trim=3cm 18.6cm 0cm 0cm,bb=96 505 532 754]{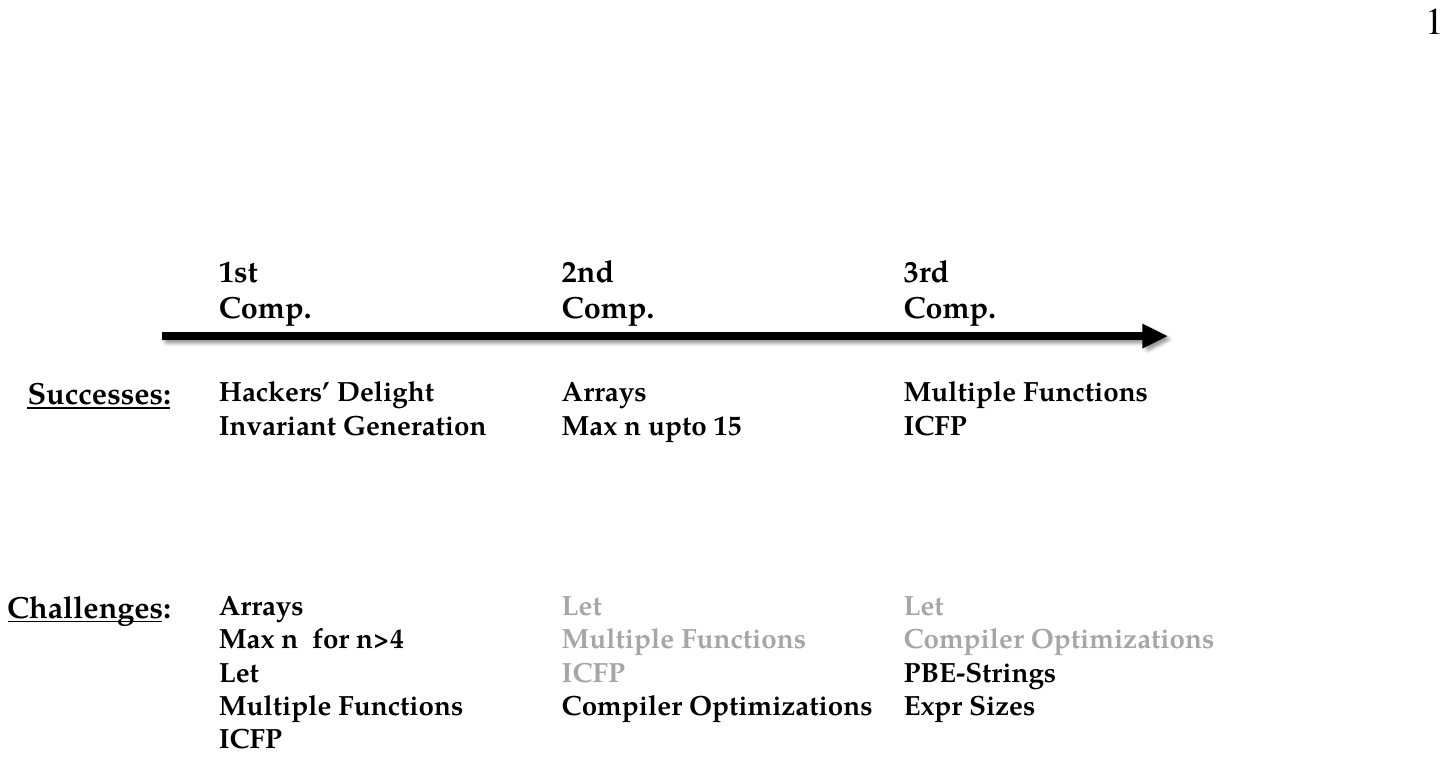}
\caption{The timeline for successful and challenging classes of benchmarks.}
\label{timelinebenchmarks}
\end{figure}

The percentage of benchmarks in the General track solved by the solvers participating in the first competition as compared to the solvers in the third competition is shown in Figure~\ref{benchmarksfirstthird}. As we can observe, a much higher fraction of benchmarks are solved by solvers in the third competition as compared to the solvers from the first competition. The successful and challenging classes of benchmarks from each competition is shown in Figure~\ref{timelinebenchmarks}. We can observe that many of the challenging benchmarks from the previous competition are tackled by the solvers in the newer competition.


\subsection*{Acknowledgments}
We would like to thank the following people for various interesting discussions related to the competition, its tracks, the SyGuS format and various other topics related to syntax-guided synthesis:
Viktor Kuncak, 
Arjun Radhakrishna,     
and
Andrew Reynolds.

We would like to thanks the StarExec~\cite{starexec} team, and especially Aaron Stump, for allowing us to use their platform and for their remarkable support for \comp's special needs.  

This research was supported by US NSF grant
CCF-1138996 (ExCAPE).

\nocite{*}
\bibliographystyle{eptcs}
\bibliography{bib-wdoi}
\end{document}